\documentclass[11pt]{article}
\usepackage{graphics}
\usepackage{graphicx}
\usepackage{epsfig}
\usepackage{amsthm,amsmath,amssymb}

\title{Extreme value analysis of actuarial risks:\\ estimation and model validation}
\author{Holger Drees\footnote{University of Hamburg, Department of Mathematics, SPST, Bundesstr.\ 55,
20146 Hamburg, Germany; email: drees@math.uni-hamburg.de}}

\newcommand{\R}{\mathbb{R}}
\newcommand{\N}{\mathbb{N}}

\def\qF{{F^\leftarrow}}

\newtheorem{theorem}{Theorem}[section]
\newtheorem{corollary}[theorem]{Corollary}
\newtheorem{remark}[theorem]{Remark}
\newtheorem{example}[theorem]{Example}

\numberwithin{equation}{section}

\newenvironment{proofof}{\noindent\sc Proof of}{
    \hspace*{\fill} $\Box$ \vspace{2ex} }

\def\eps{\varepsilon}

\def\Min(#1,#2){#1\wedge #2}
\def\Max(#1,#2){#1\vee #2}

\def\e{{\rm e}}

\def\rueck{\noindent\hangafter=1 \hangindent=1.3em}

\begin{document}

\maketitle

\begin{abstract}
  We give an overview of several aspects arising in the statistical
  analysis of extreme risks with actuarial applications in view. In particular it is demonstrated that
  empirical process theory is a very powerful tool, both for the
  asymptotic analysis of extreme value estimators and to devise
  tools for the validation of the underlying model assumptions.
  While the focus of the paper is on univariate tail risk analysis,
  the basic ideas of the analysis of the extremal dependence
  between different risks are also outlined. Here we emphasize some of the
  limitations of  classical multivariate extreme value theory and
  sketch how a different model proposed by Ledford and Tawn can help to
  avoid pitfalls. Finally, these theoretical results are used to
  analyze a data set of large claim sizes from health insurance.

\end{abstract}

\section{Introduction}

In nonlife insurance, usually extreme events constitute a
considerable portion of the total risk covered by an insurance
company. Therefore,  in actuarial practice extreme value statistics
(though often in a simplified form) has been used for at least two
decades to assess the risk of large claims. Given their exposure to
huge claims, it is natural that reinsurers were among the first to
emphasize the need for appropriate models of losses exceeding high
thresholds. While the use of Pareto distributions and
generalizations thereof were advocated early (see, e.g., Schmutz and
Doerr (1998)), the fact that they naturally arise as {\em
approximative} models for exceedances was not always fully
acknowledged, but they were often considered yet another useful
parametric model.

This situation has thoroughly changed. Nowadays it is rarely called
into question that the assessment of ``tail risks'' requires
specific methods and that extreme value theory often (though not
always) offers efficient and mathematically sound procedures to deal
with such problems. Moreover, several smooth introductions both to
general extreme value statistics and to its application to actuarial
problems have been published; see, e.g., Embrechts et al.\ (1997),
Beirlant et al.\ (2004), McNeil (1997) and Cebri\'{a}n et al.\ (2003).
For that reason, the present paper focusses on specific aspects
which have perhaps not attracted the attention they deserve:
\begin{itemize}
  \item We will show that empirical process theory offers a general
  framework to deal with different steps in the risk analysis from
  model fitting to model validation and the estimation of risk
  measures.
  \item An important step in a prudent risk assessment is to
  validate the model assumptions on which the statistical analysis
  is based. To this end, graphical tools like  qq-plots are widely
  used but the assessment which deviations from the ideal line
  indicate a violation of the model assumptions is largely
  subjective, and experience from classical statistical applications
  can be misleading if one analyzes heavy-tailed data. Hence in
  Section \ref{sect:validate} it is described how to refine such tools to obtain a
  rigorous statistical test.
  \item The analysis of the dependence between different
  extreme risks has been extensively discussed in the recent statistical
  literature. We will first comment on problems arising when
  parametric copula models are used to this end. Then we
  discuss how to overcome a serious weakness of classical
  multivariate extreme value statistics.
\end{itemize}

As the choice of topics addressed by such a partial survey is
subjective, it is inevitable that some readers will miss aspects
they consider particularly important. Perhaps the most obvious topic
we only touch on concerns the extreme value analysis of investment
risks. Although recent years have shown that in some instances the
asset side of the balance book contains the most serious risks of
extreme losses, for several reasons here we will nevertheless focus
on ``genuine'' actuarial problems related to the insured risks.
Firstly, the statistical and economic literature on extreme
investment risks is abound. Secondly, though the very basics of
extreme value theory needed in this context is the same as the one
discussed here, a serious treatment of market risks would require a
lengthy introduction to the extreme value behavior of time series;
we refrain from discussing this topic in detail in order not to
overload the article. Finally, we feel that mathematically
satisfactory solutions are yet to be developed for important
practical problems like the risk assessment for complex portfolios.
An exposition that cannot go into great details carries the risk of
provoking a serious misconception of the solutions that the state of
the art in statistical theory can actually deliver.

As there are plenty of other important problems we can merely touch
on, we will try to mitigate this lack by giving references where
aspects important in actuarial applications are discussed in greater
detail. We do however not aim at giving a full overview over the
rapidly expanding literature relevant in this context. Hence the
present text may be best characterized as a tutorial with particular
emphasis laid on crucial points which, from my personal point of
view, have often not attracted the interest they deserve.

The paper is organized as follows. Section \ref{sect:introevt} gives
an introduction into the basics of univariate extreme value theory,
with particular emphasis on conditional distributions of exceedances
(instead of the distribution of maxima as in the classical
approach). In Section \ref{sect:tailanalysis} we discuss how to
construct extreme value estimators of quantities like risk measures
and insurance premiums which depend only on the tail behavior. Then
Section  \ref{sect:validate} deals with methods to define a tail
region depending on the data and the purpose and methods of the tail
analysis as well as tools for model validation. In both these
sections, a limit theorem for the tail empirical quantile function
proves extremely useful. In Section \ref{sect:dependence} we outline
how the dependence structure between the components of a vector of
risks can be statistically analyzed. In Section
\ref{sect:dataanalysis} the previously introduced statistical
procedures are used to analyze a data set of large claims in US
health insurances. All proofs are deferred to the final Section
\ref{sect:proofs}.

\section{Basics of univariate extreme value theory}
 \label{sect:introevt}

Classically a synopsis on extreme value theory starts with the
analysis of maxima of independent and identically distributed random
variables (iid rv's). We prefer to discuss the asymptotic behavior
of excesses over high thresholds, because these naturally arise as
effective claim sizes in insurances with high retention levels,
while the maximum of claim sizes rarely is an economically
meaningful quantity.

In what follows, let $X$ denote a rv defined on some probability
space $(\Omega,\mathcal{A},P)$ with cumulative distribution function
(cdf) $F$ and quantile function  $\qF$ (i.e., the generalized
inverse of $F$). If $X$ describes a loss covered by an insurance
with retention level $u$, then
$$ F_u(x) := P(X-u\le x\mid X>u) $$
is the  cdf of the actual claim size. For a very high retention
level $u$ (e.g., in an excess of loss reinsurance against
catastrophes), usually few or none of the losses observed so far
exceed $u$, so that standard methods for risk modeling and premium
calculation do not apply directly. Of course, one could assume a
parametric model for all losses, estimate its parameters (provided
the full losses are observed) and calculate the resulting
conditional cdf of the excesses over $u$. Then, however, the fitted
model for $F_u$ is largely determined by the bulk of losses that are
much smaller than the losses of interest that exceed $u$. Hence such
an approach seems advisable only if one is confident that the same
``stochastic mechanism'' generates the  moderate losses on the one
hand and large losses on the other hand, and that all these losses
can be well described by the chosen parametric model. As this will
rarely be justified, it is widely accepted that for modeling $F_u$
one should consult only losses which are large, though perhaps still
smaller than $u$. Sometimes this general idea is subsumed in the
catchy phrase ``Let the tails speak for themselves''.

The basic idea of extreme value theory is to tackle this problem by
{\em assuming} that, after a suitable normalization, the cdf $F_u$
converges to a non-degenerate limit as the threshold $u$ tends to
the largest possible loss $\qF(1)$. More precisely, we assume that
for some (measurable) function $a>0$ there exists a non-degenerate
cdf $H$ (i.e., $H(\R)\not\subset\{0,1\}$) such that
\begin{equation}  \label{eq:excessconv}
  F_u(a(u)x) = P\Big(\frac{X-u}{a(u)}\le x \,\Big|\, X>u\Big) \to H(x)
\end{equation}
as $u\uparrow \qF(1)$ for all points of continuity $x$ of $H$ (i.e.,
$F_u(a(u)\cdot)\to H$ weakly). It turns out that then $H$ is
necessarily the cdf of a generalized Pareto distribution (GPD), that
is
$$ H(x) = H_{\gamma,\sigma} (x) = \left\{
\begin{array}{l@{\quad}c@{\quad}l}
  0 & & x\le 0,\\
  1-(1+\gamma x/\sigma)^{-1/\gamma} & \text{if} & x>0,
  1+\gamma x/\sigma>0,\\
  1 & & 1+\gamma x/\sigma\le 0, \gamma<0.
\end{array}  \right.
$$
Here $H_{0,\sigma}(x)$ is interpreted as $\lim_{\gamma\to 0}
H_{\gamma,\sigma}(x)=(1-\e^{-x/\sigma})1_{[0,\infty)}(x)$, which is
an exponential cdf. Note that the scale parameter depends on the
choice of the normalizing function $a$; we can and will always
assume $\sigma=1$ and write $H_\gamma$ instead of $H_{\gamma,1}$.

If \eqref{eq:excessconv} holds, then the conditional cdf $F_u$ can
be approximated by $H_{\gamma,\sigma_u}$ with $\sigma_u=a(u)$, {\em
provided $u$ is sufficiently large}. In that case, the following
approximation of the tail of the loss cdf $F$ follows:
\begin{equation}  \label{eq:cdfapprox}
 1-F(y) = (1-F(u))(1-F_u(y-u)) \approx
(1-F(u))\big(1-H_{\gamma,\sigma_u}(y-u)\big)
\end{equation}
for $y\ge u$.
 Of course, this approximation can also be used for thresholds $u$
different from the retention level at hand. It is important to note
that (almost) always \eqref{eq:cdfapprox} is only an {\em
approximation} to the tail and that its accuracy depends on the
choice of $u$. Hence one should avoid considering the GPD model to
be the ``true'' one above a certain threshold $u$. As we will
discuss in detail later on, there will always be a bias-variance
trade-off when choosing a threshold to estimate premiums or risk
measures.

The extreme value approach to the analysis of $F_u$ relies on
convergence \eqref{eq:excessconv}. Fortunately, almost all textbook
distributions suggested to model claim sizes fulfill this condition,
that can be reformulated as
\begin{equation}  \label{eq:necsuffcondcdf1}
  \lim_{u\uparrow \qF(1)} \frac{1-F(u+a(u)x)}{1-F(u)} =
  1-H_\gamma(x), \quad x>0,
\end{equation}
 for some $\gamma\in\R$. It is easily seen that this
 condition holds if and only if for $\tilde a(t)=a(\qF(1-t))$
\begin{equation}  \label{eq:necsuffcondqf1}
  \lim_{t\downarrow 0} \frac{\qF(1-tx)-\qF(1-t)}{\tilde a(t)} =
  H_\gamma^\leftarrow(1-x)=\frac{x^{-\gamma}-1}\gamma, \quad
  x\in(0,1),
\end{equation}
where the right-hand side is interpreted as $-\log x$ for
$\gamma=0$. (Indeed, \eqref{eq:necsuffcondqf1} holds for all $x>0$.)

The so-called {\em extreme value index} $\gamma$ largely determines
the tail behavior of $F$. If $\gamma>0$, then the loss distribution
is unbounded and \eqref{eq:necsuffcondqf1} is equivalent to the
regular variation of $1-F$ at $\infty$  and of $\qF$ at 1:
\begin{eqnarray}  \label{eq:necsuffcondcdf2}
  \lim_{u\uparrow \infty} \frac{1-F(ux)}{1-F(u)}& = &
  x^{-1/\gamma}, \\ \label{eq:necsuffcondqf2}
   \lim_{t\downarrow 0} \frac{\qF(1-tx)}{\qF(1-t)}
    & =  &   x^{-\gamma},\qquad x>0.
\end{eqnarray}
In this case, both $F$ and $X$ are called heavy-tailed. (Notice,
however, that in the literature other meanings of the term
``heavy-tailed distributions'' are common, too.) Typical examples
are Burr distributions, loggamma distributions and $t$
distributions. As the survival function $1-F(x)$ roughly decays as
the power function $x^{-1/\gamma}$, large losses are the more likely
the larger $\gamma$ is. In particular, the loss has infinite
expectation if $\gamma>1$ and it has infinite variance if $\gamma\in
(1/2,1)$.

If the extreme value index is negative, then the loss has bounded
support, while for $\gamma=0$ the right endpoint of the loss
distribution can be finite or infinite. For most textbook examples,
including lognormal, gamma and normal distributions, the latter is
true.

This article will mainly focus on the case $\gamma>0$, that is
obviously the most troublesome from an insurer's perspective. We
will see that in the statistical analysis it is nevertheless
sometimes better to work with the more general conditions
\eqref{eq:necsuffcondcdf1} and \eqref{eq:necsuffcondqf1} instead of
the simpler conditions \eqref{eq:necsuffcondcdf2} and
\eqref{eq:necsuffcondqf2}, that correspond to the particular choice
$a(u)=\gamma u$.

We close this section with a brief outline of the relationship to
the limit behavior of maxima of iid rv's $X_i$, $1\le i\le n$, with
cdf $F$. It can be shown that assumption \eqref{eq:excessconv} is
equivalent to the convergence of the suitably standardized maxima to
the (generalized) extreme value distribution corresponding to $H$,
i.e.
$$ \lim_{n\to\infty} P\Big\{\frac{\max_{1\le i\le n} X_i-b_n}{a_n}\le x\Big\}
= G(x)
$$
 holds for some $a_n>0$, $b_n\in\R$ and all points of continuity $x$ of
the non-degenerate cdf $G$ if and only if \eqref{eq:excessconv} is
fulfilled with $H=(1+\log G)^+$. Then it is said that $F$ belongs to
the maximum domain of attraction ($F\in D(G)$ for short), and one
can choose $b_n=\qF(1-1/n)$ and $a_n=a(b_n)=\tilde a(1/n)$.

\section{Univariate tail risk analysis}
\label{sect:tailanalysis}

To start with a concrete problem, assume that based on observed
losses $X_1,\ldots,X_n$ the fair net premium of a (working)
excess-of-loss (XL-) reinsurance with a cover of $c$ in excess of
 $t$ is to be estimated, that is, the reinsurer has to
pay $\min((X-t)^+,c)$ of all future claims $X$  exceeding $t$. After
a suitable correction for inflation, the random variables $X_i$,
$1\le i\le n$, shall be regarded as iid with some unknown cdf $F$.
If at most a few observations exceed the retention level $t$, then
the net premium per loss $E\big(\min((X-t)^+,c)\big)$ cannot be
 directly estimated by the corresponding mean. Therefore, we assume that $F$
fulfills the basic condition \eqref{eq:necsuffcondcdf1} for some
$\gamma\in\R$, so that we may approximate the net premium as
follows:
\begin{eqnarray}  \label{eq:netpremiumapprox}
 E\big(\min((X-t)^+,c)\big) & = & \int_t^{t+c} 1-F(s)\, ds \nonumber\\
   & = & \int_{(t-u)/a(u)}^{(t+c-u)/a(u)}
   \frac{1-F(u+a(u)x)}{1-F(u)}\, dx\cdot a(u) (1-F(u)) \nonumber\\
   & \approx & \int_{(t-u)/a(u)}^{(t+c-u)/a(u)} (1+\gamma
   x)^{-1/\gamma}\,  dx\cdot a(u) (1-F(u))\\
   & = & \bigg[
   \Big(1+\gamma\frac{t-u}{a(u)}\Big)^{1-1/\gamma}-\Big(1+\gamma\frac{t+c-u}{a(u)}\Big)^{1-1/\gamma}\bigg]
   \frac{a(u)(1-F(u))}{1-\gamma}  \nonumber
\end{eqnarray}
for some suitable threshold $u\le t$, provided that
$1+\gamma(t+c-u)/a(u)>0$. If $1-F(u)$ is sufficiently large, then it
can be estimated by the corresponding empirical probability
$1-F_n(u)$. Hence, if one replaces the extreme value $\gamma$ and
the scale factor $a(u)$ by some estimators $\hat\gamma_n$ and $\hat
a_n(u)$, respectively, then one obtains a reasonable estimator of
the net premium per claim. (To obtain an estimator for the net
premium of the whole XL reinsurance contract, one has to multiply
this expression with some estimator of the expected number of
claims.)

Estimators of $\gamma$ and $a(u)$ that use only exceedances over the
threshold $u$ can be motivated in a similar way. For example, if one
can assume that $\gamma$ is strictly positive, then by
\eqref{eq:necsuffcondcdf2} the conditional distribution of $X/u$
given $X>u$ may be approximated by a Pareto distribution with
Lebesgue density $\tilde h_\gamma(x)=x^{-(1/\gamma+1)}/\gamma$,
$x>1$. Ignoring the approximation error and the fact that the number
$N(u)$ of exceedances is also random, we may estimate $\gamma$ by a
maximum likelihood approach to obtain
\begin{equation} \label{eq:Hillestexceed}
  \hat\gamma_n := \frac 1{N(u)} \sum_{i=1}^n \log \frac{X_i}u
  1_{(u,\infty)}(X_i).
\end{equation}
If one starts with condition \eqref{eq:necsuffcondcdf1} in the
general case $\gamma\in\R$, then the conditional distribution of the
excesses $X_i-u$ given $X_i>u$ are iid with approximative density
$h_\gamma(x/\sigma)/\sigma=(1+\gamma
x/\sigma)^{-(1/\gamma+1)}/\sigma$ for $1+\gamma x/\sigma>0$ with
$\sigma:=a(u)$. As the resulting approximative likelihood is
unbounded for $\gamma\le -1$, a point of maximum of the
loglikelihood $-(1/\gamma+1)\sum_{i=1}^n
\log\big(1+\gamma(X_i-u)^+/\sigma\big)-N(u)\log \sigma$ on the
parameter set  $\{(\gamma,\sigma)\mid
\gamma>-1,\sigma>-\gamma\max_{1\le i\le n} (X_i-u)^+\}$ can be
motivated as an estimator for $(\gamma, a(u))$.

Of course, several other estimators of the extreme value index and
the scale factor have been proposed; see, for instance, de Haan and
Ferreira (2006), Sections 3 and 4.

The performance of all these estimators crucially depends on the
accuracy of the (generalized) Pareto approximation in
\eqref{eq:necsuffcondcdf1} (respectively \eqref{eq:necsuffcondcdf2}
in the case $\gamma>0$) and the choice of the threshold $u$. Too low
a threshold will lead to a large bias, because the GPD approximation
is inaccurate for the smallest exceedances of $u$. On the other
hand, if $u$ is chosen too large, then the estimators use only a
very small fraction of the whole sample and thus their variance will
be large. In Section 4, we will discuss methods to deal with the
bias-variance tradeoff in greater detail.

Because one has to choose the threshold $u$ depending on the data,
it seems natural to use a large order statistic $u=X_{n-k_n:n}$
(with $X_{i:n}$ denoting the $i$th smallest observation).  Then all
estimators under consideration are based on the $k_n+1$ largest
observations and can therefore be written as functionals of the tail
empirical quantile function
\begin{equation}  \label{eq:Qndef}
 Q_n(t) := X_{n-[k_nt]:n}, \quad t\in [0,1].
\end{equation}
For example, replacing $u$ with $X_{n-k_n:n}$ in
\eqref{eq:Hillestexceed} yields the well-known Hill estimator
\begin{equation}  \label{eq:Hilldef}
 \hat\gamma_n := \hat\gamma_{n,k_n} := \frac 1{k_n} \sum_{i=1}^{k_n} \log
\frac{X_{n-i+1:n}}{X_{n-k_n:n}}
\end{equation}
(if there are no ties).

 Since the parameters $\gamma,a(u)$ and
$\tilde a(t)$ are only defined by limit relations like
\eqref{eq:necsuffcondcdf1} and \eqref{eq:necsuffcondqf1}, the
performance of their estimators must be analyzed in an asymptotic
framework. (Indeed, there are no unique ``true'' functions $a$ and
$\tilde a$, because any function $\bar a$ such that $\bar
a(t)/\tilde a(t)\to 1$ as $t\downarrow 0$ also satisfies
\eqref{eq:necsuffcondqf1}.) Because the basic condition
\eqref{eq:necsuffcondqf1} describes the behavior of $F^\leftarrow$
only at its right endpoint $F^\leftarrow (1)$, in the asymptotic
setting we must ensure that, while the number of order statistics
used for the statistical tail analysis tends to infinity, all order
statistics tend to  $F^\leftarrow (1)$, that is, $(k_n)_{n\in\N}$ is
a so-called {\em intermediate sequence} satisfying
\begin{equation}  \label{eq:intermediatedef}
 k_n\to\infty, \quad \frac{k_n}n\to 0.
\end{equation}
Moreover, $k_n$ should not grow too fast to avoid the aforementioned
bias problems due to a poor GPD approximation. The precise
conditions on $k_n$ will be given below in terms of the
approximation error in \eqref{eq:necsuffcondqf1}, i.e.
$$ R(t,x) := \frac{\qF(1-tx)-\qF(1-t)}{\tilde a(t)}
-\frac{x^{-\gamma}-1}\gamma.
$$

 As the randomness of all
extreme value estimators under consideration is captured by the tail
empirical quantile function, it is natural first to establish a
limit theorem for this process and then to conclude the asymptotic
behavior of quite general extreme value estimators (or tests) by a
functional delta method. In what follows, we are focussing on the
case $\gamma>-1/2$ and will often assume that $\gamma\ge 0$, which
is by far the most relevant case in actuarial applications and helps
to avoid technicalities. The following limit theorem (Drees, 1998a,
Theorem 2.1) is the corner stone for the subsequent risk analysis
\begin{theorem}  \label{theo:Qnlimit}
  If $(k_n)_{n\in\N}$ is an intermediate sequence such that for
  some $\tilde\eps>0$
  \begin{equation}  \label{eq:kncond1}
     k_n^{1/2} \sup_{0<x\le 1+\tilde\eps} x^{\gamma+1/2} |R(k_n/n,x)|
     \;\to\; 0,
  \end{equation}
  then for a standard Brownian motion $W$ and all $\eps>0$ we have
  \begin{equation}  \label{eq:Qnconv}
   k_n^{1/2}  \Big( \frac{Q_n(t)-\qF(1-k_n/n)}{\tilde a(k_n/n)} -
  \frac{t^{-\gamma}-1}\gamma\Big)_{0<t\le 1} \;\longrightarrow\;
  \big(t^{-(\gamma+1)}W(t)\big)_{0<t\le 1}
  \end{equation}
  weakly in the normed vector space $\big(D_{\gamma,\eps},\|\cdot \|_{\gamma,\eps}\big)$
  of functions $z:(0,1]\to\R$
  which are continuous from the right with left-hand limits and
  finite weighted supremum norm
  $$ \|z\|_{\gamma,\eps} := \sup_{0<t\le 1} t^{\gamma+1/2+\eps}
  \|z(t) \|.
  $$
  { }
\end{theorem}
It is noteworthy that, under the basic assumption
\eqref{eq:necsuffcondqf1}, there always exist intermediate sequences
$(k_n)_{n\in\N}$ such that \eqref{eq:kncond1} is fulfilled. Hence
\eqref{eq:kncond1} is not a condition on $F$, but it merely
restricts the speed at which $k_n$ grows with the sample size.

Now suppose $T:D_{\gamma,\eps}\to\R$ is a scale and shift invariant
functional (i.e., $T(az+b)=T(z)$ for all $a>0$, $b\in\R$ and $z\in
D_{\gamma,\eps}$) such that $T(z_\gamma)=\gamma$ for
$z_\gamma(t):=(t^{-\gamma}-1)/\gamma$, such that the following
(Hadamard) differentiability condition holds: there exists a signed
measure $\nu_{T,\gamma}$ on $(0,1]$ such that
\begin{equation}  \label{eq:Tdiff}
 \frac{T(z_\gamma+\lambda_n y_n)-T(z_\gamma)}{\lambda_n}
\;\longrightarrow\; \int y\, d\nu_{T,\gamma}
\end{equation}
 for all
sequences $\lambda_n\downarrow 0$ and all $y_n,y\in D_{\gamma,\eps}$
satisfying $\|y_n-y\|_{\gamma,\eps}\to 0$. Then one may easily
deduce that
\begin{equation} \label{eq:TQnconv}
  k_n^{1/2} \big(T(Q_n)-\gamma\big) = k_n^{1/2} \Big( T\Big(
\frac{Q_n-\qF(1-k_n/n)}{\tilde a(k_n/n)}\Big)-\gamma\Big)
 \;\longrightarrow\; \int_{(0,1]} t^{-(\gamma+1)}W(t)\,
\nu_{T,\gamma}(dt) \quad \text{weakly,}
\end{equation}
where the right-hand side is normally distributed with expectation 0
and variance
\begin{equation}  \label{eq:gammaestas}
 \sigma_{T,\gamma}^2 := \int_{(0,1]}\int_{(0,1]} (st)^{-(\gamma+1)}
\min(s,t) \, \nu_{T,\gamma}(ds) \, \nu_{T,\gamma}(dt).
\end{equation}
Likewise, the scale factor $\tilde a(k_n/n)$ can be estimated by
$S(Q_n)$, where $S:D_{\gamma,\eps}\to\R$ is a scale equivariant and
shift invariant functional (i.e., $S(az+b)=aS(z)$ for all $a>0$,
$b\in\R$ and $z\in D_{\gamma,\eps}$) such that $S(z_\gamma)=1$ and
for some signed measure $\nu_{S,\gamma}$ on $(0,1]$
\begin{equation}  \label{eq:Sdiff}
\frac{S(z_\gamma+\lambda_n y_n)-S(z_\gamma)}{\lambda_n}
\;\longrightarrow\; \int y\, d\nu_{S,\gamma}
\end{equation}
for all sequences $\lambda_n\downarrow 0$ and
$\|y_n-y\|_{\gamma,\eps}\to 0$. Then we conclude
\begin{equation}  \label{eq:scaleestas}
 k_n^{1/2} \Big(\frac{S(Q_n)}{\tilde a(k_n/n)}-1\Big)
\;\longrightarrow\; \int_{(0,1]} t^{-(\gamma+1)}W(t)\,
\nu_{s,\gamma}(dt) \quad \text{weakly.}
\end{equation}
(In fact, the {\em joint} weak convergence of \eqref{eq:Qnconv},
\eqref{eq:TQnconv} and \eqref{eq:scaleestas} holds.)

As an example, consider the functional $(T(z),S(z))$ defined as a
solution $(\gamma,\sigma)$ of the equations
\begin{eqnarray*}
  \int_0^1 \frac 1{1+\gamma(z(t)-z(1))/\sigma}\, dt & = & \frac
  1{\gamma+1}\\
   \int_0^1 \log\big(1+\gamma(z(t)-z(1))/\sigma\big)\, dt & = & \gamma
\end{eqnarray*}
with $\gamma\ne 0$, or as $(0,\sigma)$ if $\big(\int_0^1 z(t)-z(1)\,
dt\big)^2 =\int_0^1 (z(t)-z(1))^2\, dt/2$ and $\sigma=\int_0^1
z(t)-z(1)\, dt$. Then $(T(Q_n),S(Q_n)$ is the aforementioned maximum
likelihood estimator in the approximating GPD model (or more
precisely, a solution of the corresponding likelihood equations).
Using the methodology sketched above, one can prove that under the
conditions of Theorem \ref{theo:Qnlimit}
$$ k_n^{1/2} {T(Q_n)-\gamma \choose S(Q_n)/\tilde a(k_n/n)-1}
\;\longrightarrow\; \mathcal{N}_{(0,\Sigma)} \quad\text{weakly with}
\quad \Sigma:=\Big( \begin{array}{cc}
  (1+\gamma)^2 & -(1+\gamma)\\ -(1+\gamma) & 2+2\gamma+\gamma^2
  \end{array} \Big).
$$
It can be shown that these estimators have minimal asymptotic
variances among all estimators $T(Q_n)$ and $S(Q_n)$ of the type
discussed above. (See Drees (1998a) and Drees et al.\ (2004) for
details.)

If $\gamma>0$, then one can always choose $\tilde a(t)=\gamma
\qF(1-t)$, so that  \eqref{eq:Qnconv} reads as
$$ k_n^{1/2} \Big(\frac{Q_n(t)}{\qF(1-k_n/n)}-t^{-\gamma}\Big)_{0<t\le 1}
\;\longrightarrow \big(\gamma t^{-(\gamma+1)}W(t)\big)_{0<t\le t}
\quad \text{weakly in } D_{\gamma,\eps}.
$$
Hence, in this case, we need not require that $T$ is shift
invariant, but merely that it is scale invariant, which allows for a
wider class of functionals. A prominent example is the Hill
estimator $T_H(Q_n)$ with $T_H(Z)=\int_0^1 \log (z(t)/z(1))\, dt$,
that is scale invariant but not shift invariant. The Hill estimator
is asymptotically normal with asymptotic variance $\gamma^2$ if
condition \eqref{eq:kncond1} is met for $\tilde
a(t)=\gamma\qF(1-t)$, which reads as
\begin{equation} \label{eq:kncond2}
  k_n^{1/2}  \sup_{0<x\le 1+\tilde\eps} x^{\gamma+1/2} \Big|
  \frac{\qF(1-k_nx/n)}{\qF(1-k_n/n)}-x^{-\gamma}\Big|
  \;\longrightarrow \; 0.
\end{equation}
Note that for some cdf's $F$ this condition imposes a much more
severe restriction on the number of order statistics used for
estimation than condition \eqref{eq:kncond1} for some other choice
of the normalizing function $\tilde a$. Thus, even if it is known
(or assumed) in advance that $\gamma>0$, it may be advisable to use
the shift invariant ML estimator in the GPD model instead of the
Hill estimator, although the latter has a smaller asymptotic
variance.
\begin{example}  \label{ex:shift}
  Assume that the following expansion of the quantile function
  holds:
  \begin{equation}  \label{eq:qfexpan}
   \qF(1-t) = d_1t^{-\gamma}+d_2+d_3t^{\rho-\gamma} +
  o(t^{\rho-\gamma})
  \end{equation}
  for some $\gamma,\rho>0$, $d_1>0$ and $d_2,d_3\ne 0$ with
  $d_2+d_3\ne 0$ if $\rho=\gamma$. Then condition
  \eqref{eq:kncond2} ensuring the asymptotic normality of the Hill
  estimator is equivalent to
  $k_n^{1/2}(k_n/n)^{\min(\gamma,\rho)}\to 0$ and hence to
  $k_n=o\big(n^{2\min(\gamma,\rho)/(2\min(\gamma,\rho)+1)}\big)$. In
  contrast, the ML estimator in the GPD model is asymptotically normal if condition \eqref{eq:kncond1}
  holds, e.g., with the choice $\tilde a(t)=\gamma d_1t^{-\gamma}$,
  which is equivalent to $k_n^{1/2}(k_n/n)^\rho\to 0$ and is hence
  fulfilled for $k_n=o(n^{2\rho/(2\rho+1)})$. For that reason, in the case $\rho>\gamma$, the
  ML estimator may use many more large order statistics than the
  Hill estimator before a significant bias shows up.
\end{example}
\begin{remark}
  It can be shown that, in the situation of Example \ref{ex:shift},
  the shift invariant estimators $T(Q_n)$ and $S(Q_n)$ are still asymptotically
  normal if $k_n\sim \lambda n^{2\rho/(2\rho+1)}$ for some
  $\lambda>0$, but then the limiting normal distribution is no
  longer centered. See Drees (1998a) or de Haan and Ferreira (2006),
  Section 3, for similar results under second order refinements of
  condition \eqref{eq:necsuffcondqf1}, that are generalizations of
  expansion \eqref{eq:qfexpan}.

  A different type of estimators for the extreme value index that
  explicitly uses these second order conditions are discussed in
  Section 4.5 of Beirlant et al.\ (2004). These estimators typically
  have a smaller bias in the more restrictive model, but their
  consistency is not ensured if only condition
  \eqref{eq:necsuffcondcdf1} or \eqref{eq:necsuffcondcdf2} is
  assumed.
\end{remark}

Next we will demonstrate by the example of an estimator for the net
premium of the XL-reinsurance discussed above that the asymptotic
normality of a huge class of extreme value estimators follows by
straightforward (though sometimes lengthy) computations. Replacing
in \eqref{eq:netpremiumapprox} the threshold $u$ with
$Q_n(1)=X_{n-k_n:n}$ and the unknown parameters by suitable
estimators, we arrive at
$$ \hat\Pi_n(t,c) := \frac{k_n}n S(Q_n) \int_{(t-Q_n(1))/S(Q_n)}^{(t+c-Q_n(1))/S(Q_n)}
(1+T(Q_n)   x)^{-1/T(Q_n)}\,  dx .
$$
As explained above, we are interested in the case that at most a few
observations exceed the retention level $t$. To reflect this crucial
feature in the asymptotic framework, one must consider a sequence of
retention levels $t=t_n$ which increases with the sample size. More
precisely, we assume
\begin{equation}  \label{eq:retentionasymp}
  \frac n{k_n} (1-F(t_n))\;\to\; 0\quad \text{and} \quad
  \frac{1-F(t_n+c_n)}{1-F(t_n)} \;\to\; \lambda\in (0,1).
\end{equation}
Despite the quite complex structure of the estimator
$\hat\Pi_n(t_n,c_n)$, its asymptotic normality follows from the
joint asymptotic normality of $T(Q_n), S(Q_n)$ and $Q_n(1)$ by
rather simple Taylor type expansions. For simplicity, we focus on
the case $\gamma\ge 0$, but an analogous result can be proved by the
same methodology for all $\gamma\in\R$. In the case $\gamma<0$,
though, the asymptotic behavior of $\hat\Pi_n$ also depends on the
asymptotics of $S(Q_n)$, while it does not play a role in the
following result.
\begin{corollary}  \label{cor:premiumestAN}
 Assume that $F\in D(G_\gamma)$ for some $\gamma\ge 0$, that
 $(t_n)_{n\in\N}$ is a sequence of retention levels  such that \eqref{eq:retentionasymp}
 holds,
 and let
 $(k_n)_{n\in\N}$ be an intermediate sequence satisfying
 \eqref{eq:kncond1},
   \begin{equation} \label{eq:kncond4}
    \sup_{0<x\le 1} x^\gamma|R(k_n/n,x)|=o(k_n^{-1/2}\tau_n)
  \end{equation}
  and
  \begin{equation} \label{eq:kncond3}
   \log\Big(\frac n{k_n}(1-F(t_n))\Big) = \left\{
   \begin{array}{l@{\quad}l}
     o(k_n^{1/2}), & \gamma>0,\\
     o(k_n^{1/6}), & \gamma=0
   \end{array} \right.
  \end{equation}
 with
   $$ \tau_n := \left\{
     \begin{array}{l@{\quad}l}
       \big|\log\big(\frac
   n{k_n}(1-F(t_n))\big)\big|/\gamma, & \gamma>0,\\
     \frac 12 \log^2\big(\frac
   n{k_n}(1-F(t_n))\big), & \gamma=0.
    \end{array}\right.
   $$
  Then
   $$ \frac n{k_n^{1/2}\tau_n\tilde a(k_n/n)}\Big(\frac
   n{k_n}(1-F(t_n))\Big)^{\gamma-1}
   \Big(\hat\Pi_n(t_n,c_n)-\int_{t_n}^{t_n+c_n} 1-F(s)\, ds\Big)
   \;\longrightarrow\; \mathcal{N}_{(0,\sigma_{\Pi,\gamma}^2)} \quad
   \text{weakly}
   $$
   with
   $$ \sigma_{\Pi,\gamma}^2 :=
   \Big(\frac{1-\lambda^{1-\lambda}}{1-\gamma}\Big)^2\sigma_{T,\gamma}^2.
   $$
\end{corollary}
\begin{remark}
  \begin{enumerate}
    \item From Corollary \ref{cor:premiumestAN} and its proof, one can easily
    conclude that
    $$ \frac{k_n^{1/2}}{\tau_n} \Bigg( \frac{\int_{t_n}^{t_n+c_n}
    1-F(s)\, ds}{\hat\Pi_n(t_n,c_n)}-1\Bigg)\;\longrightarrow\;
    \mathcal{N}_{(0,\sigma_{T,\gamma}^2)} \quad \text{weakly}.
    $$
    Hence, if $\hat\tau_n$ is a consistent estimator of $\tau_n$ in
    the sense that $\hat\tau_n/\tau_n\to 1$ in probability, and if
    $\sigma_{T,\gamma}^2$ is a continuous function of $\gamma$, then
    $$ \frac{k_n^{1/2}}{\hat\tau_n\sigma_{T,T(Q_n)}} \Bigg( \frac{\int_{t_n}^{t_n+c_n}
    1-F(s)\, ds}{\hat\Pi_n(t_n,c_n)}-1\Bigg)\;\longrightarrow\;
    \mathcal{N}_{(0,1)} \quad \text{weakly},
    $$
    from which asymptotic confidence intervals are readily
    constructed. If $\gamma>0$, then
    $$ \hat\tau_n := \frac{\log(t_n/Q_n(1))}{T^2(Q_n)} $$
    is a consistent estimator of $\tau_n$.
    \item In the situation of Example \ref{ex:shift}, condition
    \eqref{eq:kncond4} is a direct consequence of condition
    \eqref{eq:kncond1}. In general, though, \eqref{eq:kncond4}
    cannot always be fulfilled, if the rate of convergence in
    \eqref{eq:necsuffcondqf1} is particularly slow. Since in the proof of Corollary \ref{cor:premiumestAN} this
    condition is essentially only needed to bound the bias term IV,
    a closer inspection of the proof shows that it can be replaced
    by a weaker, but more complex condition on $k_n$ which can
    always be satisfied.
  \end{enumerate}
\end{remark}

By the same approach one can construct estimators of all risk
measures or insurance premiums which are smooth functionals of the
tail cdf $1-F(t)$, $t>u$, for some large $u$ or of the tail quantile
function $\qF(1-t)$, $t<\eta$, for some small $\eta>0$. For example,
the value at risk $\qF(1-\alpha)$ for small $\alpha$ can be
estimated by
$$ \widehat{VaR_\alpha} := Q_n(1)+S(Q_n)
\frac{(n\alpha/k_n)^{-T(Q_n)}-1}{T(Q_n)}
$$
(cf.\ Drees (2003)). Extreme value estimators of reinsurance
premiums according to Wang's premium principle have been examined by
Vandewalle and Beirlant (2006) in the case $\gamma>0$ (without using
the tail empirical quantile function explicitly).

An advantage of the approach via the tail empirical quantile
function is that, with rather little effort, one can analyze the
asymptotic behavior of a large class of extreme value estimators in
a unified framework, and hence easily compare their performance.
Moreover, the same analysis immediately  gives the asymptotic
normality of the estimators if one replaces the assumption of
independence of the observations by more general condition on the
serial dependence structure. Indeed, Drees (2003) proved the
convergence of the tail empirical quantile function $Q_n$ towards a
centered Gaussian process for stationary time series which satisfy
suitable mixing conditions. Although all the estimators discussed
above can still be used in this more general setting, usually their
estimation error will be larger than for iid data. In an extensive
simulation study, Drees (2003) showed that then the actual coverage
probability of confidence intervals for extreme quantiles
constructed on the basis of the theory for iid data can be much
smaller than its nominal value. Therefore, it is important not to
use these confidence intervals when analyzing time series of returns
on some investment that usually exhibit quite strong a serial
dependence.

\section{Selecting the tail fraction and validating the model}
  \label{sect:validate}

As explained above, in almost all cases there does not exist a
threshold $u$ such that the tail cdf $F(x), x>u$, is {\em exactly}
equal to some GPD tail, but the accuracy of the GPD approximation
usually increases with the threshold. Consequently, roughly speaking
the modulus of the bias of any of the extreme value estimators
discussed in  Section \ref{sect:tailanalysis} will be a
monotonically decreasing function of $u$, respectively an increasing
function of $k$ if the order statistic $X_{n-k:n}$ is used as the
threshold. (This statement should be taken with a pinch of salt: for
very small $k$ the bias sometimes becomes larger again, but in an
asymptotic setting the monotonicity can be made precise for
intermediate sequences $(k_n)_{n\in\N}$.) On the other hand, the
variance is an increasing function of $u$ and decreasing function of
$k$, respectively. Therefore, choosing an ``optimal'' sample
fraction of largest observations used for the statistical tail
analysis involves a bias-variance tradeoff. Note that this selection
does not only depend on the data set (or the underlying
distribution), but also on the estimator (or statistical test) used
in the analysis. Moreover, the appropriate balance between bias and
 variance may also depend on the purpose of the statistical
analysis: in some applications a non-negligible bias may be
unacceptable when calculating an insurance premium, while such a
bias may be admissible if it helps to reduce the variance of an
estimator of a risk measure. Thus, for a given data set, there does
not exist {\em the} optimal choice for $u$ or $k$.

This said, widely applicable techniques are needed to select the
number of largest order statistics used in the statistical analysis.
The most popular graphical tool is to plot the estimator under
consideration (based on the largest $k+1$ observations) versus $k$.
Typically, the graph will be rather wiggly for small values of $k$,
and it will be more or less monotone for large values of $k$ due to
the increasing modulus of the bias. Hopefully, there is a range in
between where the plot is relatively stable, indicating that the
bias is not yet dominating, but the variance has already decreased
to an acceptable level. Drees et al.\ (2000) showed (for the Hill
estimator) that it may be advisable to plot the estimator versus
$\log k/\log n$ (as suggested by C.\ St\u{a}ric\u{a}), because
usually this graph spends a larger portion of time in the
neighborhood of the true value.

\begin{figure}[htb]
\centerline{\includegraphics[width=140mm]{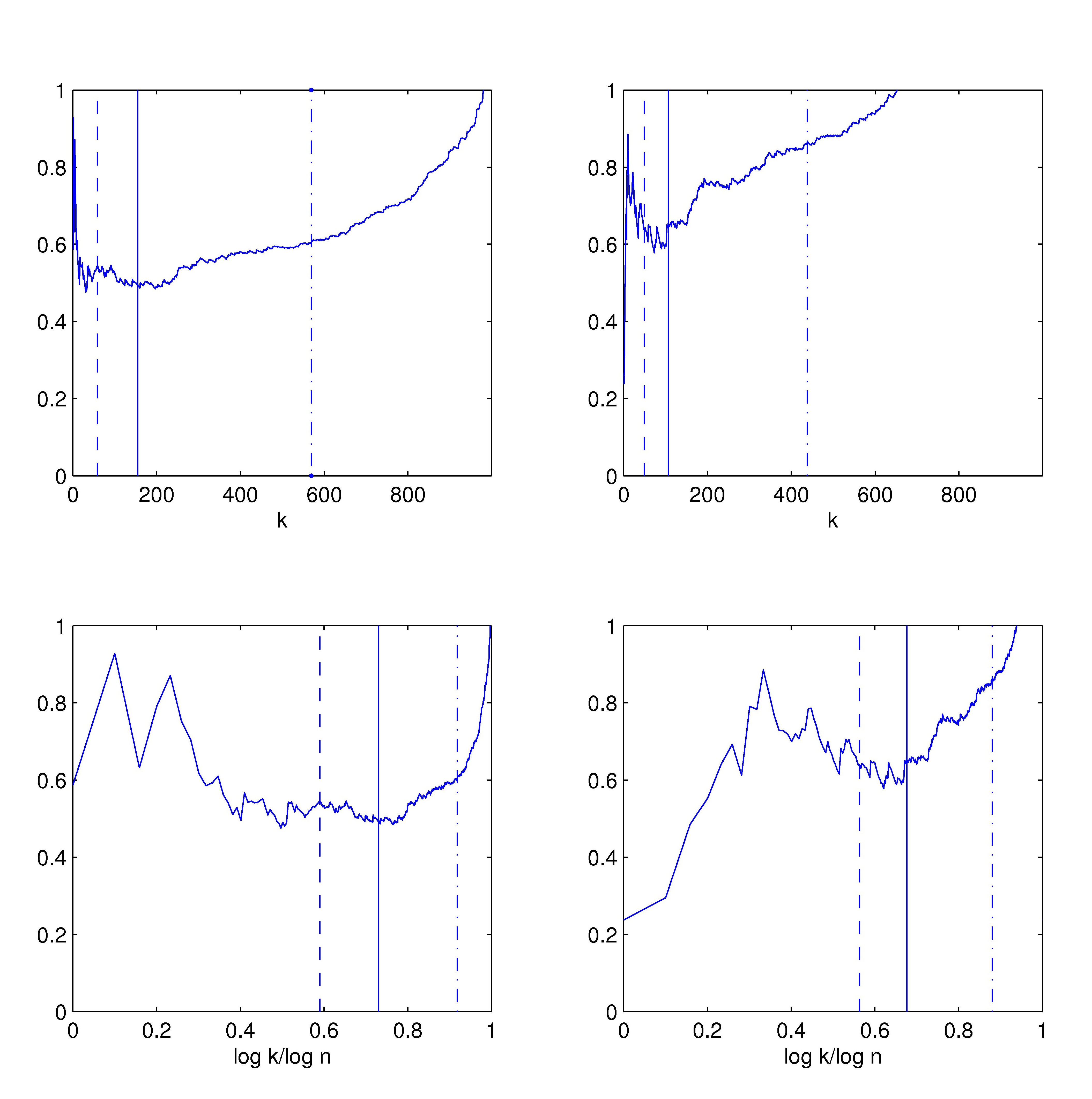}}
\vspace*{-0.4cm} \caption{Hill estimator based on $k+1$ largest
order statistics versus $k$ (above) and versus $\log k/\log n$
(below) for $n=1000$ iid Fr\'{e}chet rv's (left) and logarithmically
disturbed Pareto rv's (right) with $\gamma=1/2$. Estimated optimal
numbers $\hat k_n^{boot}$ (dashed), $\hat k_n^{seq}$ (solid) and
$\hat k_n^{ML}$ (dash-dotted) are indicated by vertical lines.}
 \label{fig:Hillplots}
\end{figure}

Figure \ref{fig:Hillplots} shows such plots for the Hill estimator
calculated for a sample of $n=1000$ iid Fr\'{e}chet rv's with cdf
$F(x)=\exp(-x^{-1/\gamma})$ on the left-hand side and for a sample
of $n$ rv's with quantile function $\qF(1-t)=(t/|\log t|)^{-\gamma}$
on the right-hand side with $\gamma=1/2$. The plots for the Fr\'{e}chet
rv's are quite stable for $k$ around 150 or $\log k/\log n$ about
0.7. In contrast, in the right-hand plots for the logarithmically
disturbed Pareto distribution, after strong fluctuations in the
beginning, the graphs immediately show a clear upward trend, and so
no plateau is clearly visible. This different behavior is caused by
the different accuracy of the GPD approximation to the tail. While
the Fr\'{e}chet distribution satisfies expansion \eqref{eq:qfexpan} with
$\rho=1$ leading to the optimal rate of convergence when $k_n$ is of
the order $n^{2/3}$, in the case of the logarithmically disturbed
Pareto distribution it can be shown that the squared bias dominates
the variance if $k_n$ is of larger order than $\log^2 n$. Thus for
the second distribution the increasing bias leads to the clear trend
already for quite small values of $k$.

To give some advice how to choose the sample fraction used in the
tail analysis in such cases and also to avoid subjective choices
which have an influence on the estimation accuracy that is difficult
to quantify, fully automatic data-driven selection procedures have
been proposed that minimize the asymptotic mean squared error of the
estimators under consideration.
 Here we consider three different methods to choose the number
 of largest order statistics such that the (asymptotic) mean squared
 error of the Hill estimator $\gamma_{n,k}$ is minimized. See
 Section 4.7 of Beirlant et al.\ (2004) for a more extensive list and
 additional references.

Danielsson et al.\ (2001) used a bootstrap approach to minimize the
mean squared error (MSE). Hill (1990) showed that the standard
bootstrap does not work here, because it does not capture the bias
of the Hill estimator (and other linear statistics) properly, but a
suitable bootstrap may yield a consistent estimator of the MSE in a
restricted model. Instead of trying to minimize the MSE of the Hill
estimator directly, Danielsson et al.\ (2001) used the auxiliary
statistic $A_{n,k} := (M_{n,k}-2\hat\gamma_{n,k}^2)^2$ with
\begin{equation} \label{eq:Mnkdef}
    M_{n,k} := k^{-1} \sum_{i=1}^k \log ^2 (X_{n-i+1:n}/X_{n-k:n}).
\end{equation}
 It can be shown that under a suitable second order condition,
that generalizes assumption \eqref{eq:qfexpan}, a sequence $\bar
k_0(n)$ which minimizes $E(A_{n,k})$ and a sequence $k_0(n)$ which
minimizes the MSE of the Hill estimator $\hat\gamma_{n,k}$ have the
same asymptotic behavior up to a multiplicative constant. Starting
from this fact, Danielsson et al.\ (2001) developed the following
algorithm:
\begin{itemize}
  \item For some $\eps\in(0,1/2)$, some $n_1=O(n^{1-\eps})$ and
  $n_2:=[n_1^2/n]$, define $k_{0}^*(n_i)$, $i=1,2$, which
  minimizes the conditional expectation of $(M_{n,k}^*-2(\hat\gamma_{n,k}^*)^2)^2$
  given the data $X_1,\ldots, X_n$, where $\hat\gamma_{n,k}^*$ and
  $M_{n,k}^*$ are defined as in \eqref{eq:Hilldef} and \eqref{eq:Mnkdef},
  respectively, but with $n$ replaced by $n_1$ and $X_i$ replaced by $X_i^*$ independently
  drawn from $X_1,\ldots,X_n$ (with replacement). Here the
  conditional expectation is minimized over $k_1\in\{[\log
  n_1],\ldots, [n_1/\log n_1]\}$, say.
  \item The asymptotic MSE of the Hill estimator $\hat\gamma_{n,k}$
  is then minimal for
  $$ \hat k_n^{boot} := \frac{(k_{0}^*(n_1))^2}{k_{0}^*(n_2)}
  \big(2\log n_1/\log k_{0}^*(n_1)-1\big)^{2(\log
  k_{0}^*(n_1)/\log n_1-1)}.
  $$
\end{itemize}
The performance of the data-driven choice of $k$ crucially depends
on the value $n_1$. Using heuristic arguments, Danielsson et al.\
(2001) proposed to select $n_1$ which minimizes
$(Q(n_1,k_{0}^*(n_1)))^2/$ $Q(n_2,k_{0}^*(n_2))$ with $Q(n_i,k_i)$
denoting the conditional expectation of
$\big(M_{n_i,k_i}^*-2(\hat\gamma_{n_i,k_i}^*)^2\big)^2$ given the
data.

 Drees and Kaufmann (1998) suggested a
sequential procedure that was inspired by the so-called
Lepskii-method for adaptive bandwidth selection in curve estimation.
The basic idea of this approach is that too large a difference
between two Hill estimators $\hat\gamma_{n,i}$ and
$\hat\gamma_{n,k}$ with $i<k$ indicates that the latter exhibits a
large bias. As the random error of the difference is of the order
$i^{-1/2}$, an asymptotically optimal choice of the number of order
statistics can be determined from the smallest $k$ such that
$i^{1/2}|\hat\gamma_{n,i}-\hat\gamma_{n,k}|$ exceeds a suitable
threshold. More precisely, Drees and Kaufmann (1998) proposed the
following algorithm:
\begin{itemize}
  \item For some $r_n=o(n^{1/2})$ such  that $(\log\log
  n)^{1/2}=o(r_n)$ let
  $$\bar k_n(r_n):= \min \big\{
  k\in\{1,\ldots,n\} \mid \max_{1\le i\le k}
  i^{1/2}|\hat\gamma_{n,i}-\hat\gamma_{n,k}|>r_n\big\}.
  $$
  \item Fix some $\lambda,\xi\in (0,1)$ such that $(\log\log
  n)^{1/(2\xi)}=o(r_n)$ and calculate a pilot estimate $\hat\gamma_n=\hat\gamma_{n,[2\sqrt{n^+}]}$
  with $n^+$ denoting the number of positive observations. Then the asymptotic MSE of the Hill
  estimator $\hat\gamma_{n,k}$ is minimal for
  $$ \hat k_n^{seq} := \bigg[ (2\hat\rho_n+1)^{-1/\hat\rho_n}
  (2\hat\gamma_n\hat\rho_n)^{1/(2\hat\rho_n+1)} \Big(\frac{\bar
  k_n(r_n^\xi)}{(\bar k_n(r_n))^\xi}\Big)^{1/(1-\xi)}\bigg]
  $$
  with
  $$ \hat\rho_n := \log \frac{\max_{1\le i\le [\lambda\bar
  k_n(r_n)]} i^{1/2}\big|\hat\gamma_{n,i}-\hat\gamma_{n,[\lambda
  \bar k_n(r_n)]}\big|}{\max_{1\le i\le \bar
  k_n(r_n)} i^{1/2}\big|\hat\gamma_{n,i}-\hat\gamma_{n,
  \bar k_n(r_n)}\big|} \Big/ \log \lambda-\frac 12.
  $$
\end{itemize}
Here the specific values of $r_n$, $\xi$ and (to a lesser extent)
$\lambda$ influence the performance of the procedure. the authors
recommended to choose $r_n=2.5 \hat\gamma_n n^{1/4}$, $\xi=0.7$ and
$\lambda=0.8$; see Drees and Kaufmann (1998) for further comments on
the implementation of this algorithm.

In yet another approach, Beirlant et al.\ (2004), Section 4.7.1
(ii), fitted an extended Pareto model with an explicit second order
correction term to the data using a maximum likelihood estimator.
Then they calculated and minimized the MSE of the Hill estimator
directly from this fit. The resulting estimated optimal number will
be denoted by $\hat k_n^{ML}$.

In Figure \ref{fig:Hillplots} the resulting estimates for the
optimal number of order statistics are indicated by vertical lines.
While the bootstrap (dashed line) and the sequential approach (solid
line) both yield reasonable values, the method which uses an
explicit model for the second order term leads to too large values
and thus a considerable bias. Indeed, it has been observed in
literature that for moderate sample sizes it is notoriously
difficult to estimate the second order parameters like $\rho$ in
expansion \eqref{eq:qfexpan}. In contrast, the bootstrap and the
sequential approach both use estimates of this second order
parameter only in an estimate of a multiplicative constant, while
their order of magnitude does not explicitly depend on such an
estimate. Hence they yield reasonable estimates even if this second
order parameter is fixed (e.g., $\rho=1$) and misspecified.

Once the sample fraction of largest order statistics has been
chosen, one should check whether it can be well approximated by a
(generalized) Pareto distribution. A classical graphical tool for
such an model validation is the qq-plot. If $F\in D(G_\gamma)$ for
some $\gamma>0$ and $k$ is chosen not too large, then using
\eqref{eq:necsuffcondqf2} one can approximate
$$ \log\frac{X_{n-i+1:n}}{X_{n-k:n}} \approx \log
\frac{\qF(1-(i-1/2)/n)}{\qF(1-(k+1/2)/n)} \approx -\gamma \log
\frac{i-1/2}{k+1/2}.
$$
Hence the points
$\big(-\log((i-1/2)/(k+1/2)),\log(X_{n-i+1:n}/X_{n-k:n})\big)$
should approximately lie on the line with slope
$\hat\gamma_n=T(Q_n)$ through the origin. To assess whether the
observed  deviations of the points from this line are probably only
due to their randomness or whether they indicate that the GPD
approximation is inaccurate, we can again use Theorem
\ref{theo:Qnlimit}.

\begin{corollary} \label{cor:qqapprox}
  Assume that $(k_n)_{n\in\N}$ is an intermediate sequence such that
  condition \eqref{eq:kncond2} holds for some  $\gamma>0$.
  Then for all $\eps>0$ and all scale invariant functionals $T$ on $D_{\gamma,\eps}$ satisfying
  $T(z_\gamma)=\gamma$ and the differentiability condition
  \eqref{eq:Tdiff}, we have
  \begin{eqnarray} \label{eq:qqapprox0}
    \lefteqn{k_n^{1/2} \Big( \log \frac{Q_n(t)}{Q_n(1)} + T(Q_n)\log
  t\Big)_{0<t\le 1}}\nonumber\\
   &\longrightarrow & \Big(
  \gamma(t^{-1}W(t)-W(1)) + \int_{(0,1]}
  s^{-(\gamma+1)}W(s)\,\nu_{T,\gamma}(ds)\cdot\log t\Big)_{0<t\le 1}
  \end{eqnarray}
  weakly in $\big(D_{0,\eps},\|\cdot\|_{0,\eps}\big)$.
  Hence,
  \begin{eqnarray} \label{eq:qqapprox1}
    \lefteqn{P\Big\{ \max_{1\le i\le k_n} h\Big(\frac{i-1/2}{k_n+1/2}\big)\Big|
    \log \frac{X_{n-i+1:n}}{X_{n-k_n:n}} + T(Q_n)\log
    \frac{i-1/2}{k_n+1/2}\Big|>c\Big\}} \nonumber \\
    & \longrightarrow &
    P\Big\{ \sup_{0<t\le 1} h(t) \Big| \gamma(t^{-1}W(t)-W(1)) + \int_{(0,1]}
  s^{-(\gamma+1)}W(s)\,\nu_{T,\gamma}(ds)\cdot\log t\Big|>c\Big\}
  \end{eqnarray}
  for all continuous functions $h:(0,1]\to(0,\infty)$ such that
  $h(t)t^{-(1/2+\eps)}\to 0$ as $t\downarrow 0$ for some $\eps>0$.
  In particular, for $T=T_H$ (i.e., $T(Q_n)$ equal to the Hill
  estimator), we obtain
  \begin{eqnarray} \label{eq:qqapprox2}
    \lefteqn{P\Big\{ \max_{1\le i\le k_n} h\Big(\frac{i-1/2}{k_n+1/2}\big)\Big|
    \log \frac{X_{n-i+1:n}}{X_{n-k_n:n}} + T_H(Q_n)\log
    \frac{i-1/2}{k_n+1/2}\Big|>cT_H(Q_n)\Big\} } \nonumber\\
    & \longrightarrow &
    P\Big\{ \sup_{0<t\le 1} h(t) \Big| t^{-1}W(t)-W(1) + \int_{(0,1]}
  s^{-1}W(s)-W(1)\,ds\cdot\log t\Big|>c\Big\}.
  \end{eqnarray}
\end{corollary}
Under slightly different conditions, a similar result has been
proved by Dietrich et al.\ (2002) for the Hill estimator.

Using \eqref{eq:qqapprox2} one can turn the Pareto qq-plot into a
statistical testing tool with given asymptotic size $\alpha$. To
this end, for some function $h$ satisfying the conditions of
Corollary \ref{cor:qqapprox}, using Monte Carlo simulations, one
determines a critical value $c_\alpha$ such that the probability on
the right-hand side of \eqref{eq:qqapprox2} equals $\alpha$. Then
with probability of about $1-\alpha$ all points of the Pareto
qq-plot should lie in the band defined by the graphs of the
functions $T_H(Q_n)(\log t\pm c_\alpha/h(t))$ if the GPD
approximation is accurate enough so that the bias of the Hill
estimator is negligible. The choice of the function $h$ determines
in which part of the qq-plot deviations are most easily detected:
the larger $h(t)$ is, the more narrow is the band at that point.
Because of the condition $h(t)t^{-(1/2+\eps)}\to 0$ as $t\downarrow
0$, the band always widens for small values of $t$, thus allowing
for larger deviations of the most extreme points of the qq-plot from
the ideal line.

It has been suggested to use such tests also to select the tail
fraction to be analyzed by increasing $k$ until the test rejects the
GPD hypothesis (see, e.g., Dietrich et al., 2002, Remark 2). In some
applications this approach may be problematic if $\alpha$ is chosen
as small as it is common in testing (e.g., $\alpha=0.05$). Note that
the limiting Gaussian process in \eqref{eq:qqapprox0} tends to 0 as
$t$ tends to 1, but that the function $h$ is assumed continuous and
hence bounded, so that deviations of points of the qq-plot from the
ideal line near $t=1$ (corresponding to the smallest order
statistics taken into account) are usually difficult to detect.
Hence it may happen that one increases the number $k$ so much that
the Hill estimator (and other extreme value estimators) are strongly
biased, before the tests acknowledges that the last order statistics
taken into account are poorly fitted. (The same argument also
applies if $L_2$-type tests like the one examined by Dietrich et
al.\ (2002) are used.)

To avoid such effects, one might think of choosing a weight function
$h$ that tends to $\infty$ as $t$ tends to 1 to compensate for the
decrease of the modulus of the limiting Gaussian process. For
instance, as this process has the variance function
$\sigma^2(t)=t^{-1}-1-\log^2 t$ if $T=T_H$, one might be tempted to
use a weight function of the form $h(t)=(t(1-t))^\eps/\sigma(t)$ for
some small $\eps>0$. Unfortunately, without additional conditions on
the smoothness of $\qF$, convergence \eqref{eq:qqapprox2} need not
hold for such a choice of $h$. The reason is that under the general
condition \eqref{eq:necsuffcondcdf2} small jumps of $\qF$ (or
continuous small but rapid changes) are still possible which lead to
an ``unusually'' irregular behavior of the tail empirical quantile
function $Q_n$ near 1. If one strengthens condition
\eqref{eq:necsuffcondcdf2} to a regularity condition on the quantile
density function $(\qF)'(1-t)=f(\qF(1-t))$, though, then assertion
\eqref{eq:qqapprox0} may also be strengthened. It is well known that
if $\qF$ has a Lebesgue density which is monotonically decreasing in
a neighborhood of 1, then by Karamata's theorem
$$ \eta(t) := \frac{tf(\qF(1-t))}{\qF(1-t)} -\gamma\;\to\; 0
$$
(cf.\ de Haan and Ferreira (2006), Proposition B.1.9 11.). If we
replace condition \eqref{eq:kncond2} with a condition on $k_n$ in
terms of the function $\eta$, then convergence \eqref{eq:qqapprox0}
can be made more precise in a neighborhood of $t=1$.
\begin{corollary} \label{cor:qqapprox1}
  Assume that  $(k_n)_{n\in\N}$ is an intermediate sequence such
  that  for some  $\gamma>0$
  \begin{equation} \label{eq:kncond6}
    k_n^{1/2} \sup_{0< t\le (1+\tilde\eps)k_n/n} |\eta(t)|\;\to\; 0
  \end{equation}
  and that the functional $T$ satisfies the conditions specified in Corollary
  \ref{cor:qqapprox}.  Then for all $\eps>0$ and  for all continuous functions $h:(0,1)\to(0,\infty)$ such that
  $h(t)t^{-(1/2+\eps)}\to 0$ as $t\downarrow 0$,  and
  $h(t)(1-t)^{1/2-\eps}\to 0$ as $t\uparrow 1$, one has
   \begin{eqnarray} \label{eq:qqapprox4}
    \lefteqn{k_n^{1/2} \bigg(h(t)\Big( \log \frac{Q_n(t)}{Q_n(1)} + T(Q_n)\log
  t\Big)1_{(0,1-1/(2k_n)]}(t)\bigg)_{0<t<1}}\nonumber\\
   &\longrightarrow & \bigg(
  h(t) \Big(\gamma(t^{-1}W(t)-W(1)) + \int_{(0,1]}
  s^{-(\gamma+1)}W(s)\,\nu_{T,\gamma}(ds)\cdot\log t\Big)\bigg)_{0<t< 1}
  \end{eqnarray}
  weakly with respect to the supremum norm.
  In particular, convergence \eqref{eq:qqapprox2}
   holds  for $T=T_H$.
\end{corollary}
In Section \ref{sect:dataanalysis} this result is applied to
construct a ``confidence band'' for the Hill-qq-plot based on claim
sizes from health insurances.

If one uses some estimator of $\gamma$ different from the Hill
estimator, then usually the probability on the right-hand side of
\eqref{eq:qqapprox1} is a continuous non-linear function of the
unknown extreme value index $\gamma$. In that case, one can still
construct tests with prescribed asymptotic size $\alpha$ for the
null hypothesis that \eqref{eq:kncond2} holds. To this end, one
first estimates $\gamma$ consistently by $\hat\gamma_n=T(Q_n)$ and
then, using Monte Carlo simulations, one determines a critical value
$c_\alpha$ such that the right-hand side of \eqref{eq:qqapprox1}
equals $\alpha$ when $\gamma$ is replaced with $\hat\gamma_n$.

Since the supremum is difficult to simulate, it seems natural to
simulate the limiting process $Z(t)= t^{-1}W(t)-W(1) +\int_{(0,1]}
s^{-(\gamma+1)} W(s)\,\nu_{T,\gamma}(ds)\cdot\log t$ on a fine grid
$t_i, 1\le i\le m$, and then to approximate the supremum by
$\max_{1\le i\le m} h(t_i) |Z(t_i)|$. This can still be
computationally
 challenging if one tries to approximate the integral of $s^{-(\gamma+1)}
 W(s)$ using some quadrature formula, because the integrand is
 unbounded in a neighborhood of zero and a large number of
 integration points may be needed to obtain an accurate
 approximation. Fortunately, for most estimators, one can avoid numerical integration
 using the fact that conditionally on $W(t_i)$, $1\le i\le m$, the
 integrals $\int_{(t_{i-1},t_i]} s^{-(\gamma+1)}
 W(s)\,\nu_{T,\gamma}(ds)$, $1\le i\le m$, (with $t_0:=0$) are independent normal
 rv's with mean
 \begin{eqnarray*}
  \mu_i & := & \int_{(t_{i-1},t_i]} s^{-\gamma} \,\nu_{T,\gamma}(ds)
 \frac{W(t_i)-W(t_{i-1})}{t_i-t_{i-1}} \\
  & & \hspace{1cm} + \int_{(t_{i-1},t_i]} s^{-(\gamma+1)} \,\nu_{T,\gamma}(ds)
 \Big(W(t_{i-1})-
 \frac{t_{i-1}}{t_i-t_{i-1}}(W(t_i)-W(t_{i-1}))\Big)
 \end{eqnarray*}
 and variance
 $$ \sigma_i^2 := \frac 1{t_i-t_{i-1}} \int_{(t_{i-1},t_i]}  \int_{(t_{i-1},t_i]}
 (st)^{-(\gamma+1)}(\min(s,t)-t_{i-1})(t_i-\max(s,t)) \,\nu_{T,\gamma}(ds)
 \,\nu_{T,\gamma}(dt).
 $$
 This statement, in turn, follows from the conditional independence
 of the processes $(W(t))_{t_{i-1}\le t\le t_i}$ which have the same
 conditional distribution as
 $$ W(t_{i-1}) - \frac{t_{i-1}}{t_i-t_{i-1}}(W(t_i)-W(t_{i-1})) +
 \tilde W(t-t_{i-1})-\frac{t-t_{i-1}}{t_i-t_{i-1}}\tilde W(t_i-t_{i-1})
 $$
 with $\tilde W$ denoting a Brownian motion independent of $W$.
 Hence the limiting Gaussian process $Z$ can be simulated as
 follows if integrals of power functions with respect to $\nu_{T,\gamma}$ can be calculated analytically:
 \begin{enumerate}
   \item Simulate $m$ independent centered normal rv's $\Delta_i$
   with variance $t_i-t_{i-1}$ where $t_0:=0$ and $t_m:=1$, and let
   $W(t_l):=\sum_{i=1}^l \Delta_i$.
   \item Simulate a normal rv $I$ with mean $\sum_{i=1}^m \mu_i$ and
   variance $\sum_{i=1}^m \sigma_i^2$.
   \item Then $\big(\gamma(t_i^{-1}W(t_i)-W(t_m))+I\log
   t_i)\big)_{1\le i\le m}$ is a (pseudo) realization of $(Z(t_i))_{1\le i\le
   m}$.
 \end{enumerate}
 Note that the variance of $I$ does not depend on $W$. In case of
 the Hill estimator and equidistant design points $t_i=i/m$, one has
 \begin{eqnarray*}
  \sum_{i=1}^m \mu_i & = & (1+\log m)\Delta_i + \sum_{i=2}^m\Big(\log
 \frac mi+1-(i-1)\log\frac i{i-1}\Big)\Delta_i\\
  \sum_{i=1}^m \sigma_i^2 & = & 1- \frac 1m \sum_{i=1}^{m-1}
  i(i+1)\log^2\frac{i+1}i.
 \end{eqnarray*}

 The Corollaries \ref{cor:qqapprox} and \ref{cor:qqapprox1} describe
 tests for the null hypothesis that the left-hand sides of \eqref{eq:kncond2} and
 \eqref{eq:kncond6}, respectively, are negligible. Likewise, one can
 devise analogous tests for the validity of condition
 \eqref{eq:kncond1}, but the resulting limiting process is more
 complicated.
\begin{corollary} \label{cor:qqapprox2}
  Suppose that $(k_n)_{n\in\N}$ is an intermediate sequence
  satisfying  condition \eqref{eq:kncond1} for some $\gamma\ge 0$, and that $S,T:D_{\gamma,\eps}\to\R$
  are scale and location invariant functionals such that $S(z_\gamma)=1$, $T(z_\gamma)= \gamma$ and
  the differentiability conditions \eqref{eq:Tdiff} and \eqref{eq:Sdiff} are met.
  Then for all $\eps>0$
  \begin{eqnarray*} \label{eq:qqapprox5}
    \lefteqn{k_n^{1/2} \Big( \log \frac{Q_n(t)-Q_n(1)}{S(Q_n)} + -\frac{t^{-T(Q_n)}-1}{T(Q_n)}
    \Big)_{0<t\le 1} \;\longrightarrow\; \Big(
 t^{-( \gamma+1)}W(t)-W(1)} \\
   & & { } -
 \int_{(0,1]}  s^{-(\gamma+1)}W(s)\,\nu_{T,\gamma}(ds)\cdot\frac{1-t^{-\gamma}(1+\gamma\log t)}{\gamma^2}
 -\int_{(0,1]}  s^{-(\gamma+1)}W(s)\,\nu_{S,\gamma}(ds)\cdot\frac{t^{-\gamma}-1}\gamma\Big)_{0<t\le
 1} \nonumber
  \end{eqnarray*}
  weakly in $\big(D_{\gamma,\eps},\|\cdot\|_{\gamma,\eps}\big)$ (with $(1-t^{-\gamma}(1+\gamma\log t))/\gamma^2$
  interpreted as $(\log^2t)/2$ for $\gamma=0$).
\end{corollary}
The proof is omitted as the assertion follows readily from Theorem
\ref{theo:Qnlimit}, \eqref{eq:gammaestas},  \eqref{eq:scaleestas}
and a Taylor expansion of $\gamma\mapsto (t^{-\gamma}-1)/\gamma$.

Similarly as above, from Corollary \ref{cor:qqapprox2} one may
construct bands around the fitted generalized Pareto quantile
function $Q_n(1)+S(Q_n)(t^{-T(Q_n)}-1)/T(Q_n)$ in which all the
points $\big((i-1/2)/(k_n+1/2),X_{n-i+1:n}\big)$ should lie with
probability of about $1-\alpha$.

\section{Analysis of the extremal dependence}  \label{sect:dependence}

In recent years, the analysis of the dependence between the extremes
of the components of a random vector of risks has attracted much
attention. If the random vector describes returns on different
assets, then it is obviously important to assess the risk of large
losses in different assets at the same time. However, extremal
dependence also matters in the analysis of claim sizes from one
customer in different lines of business. For example, if both a
building and its content are insured with the same company, then a
fire will often lead to large claims in both lines of business.
Likewise, some dependence can be expected between the claims in
different types of health insurances (e.g.\, inpatient and
outpatient cover); cf.\ Section \ref{sect:dataanalysis}.

In order not to overload this presentation, we will only sketch the
basic theory used in the analysis of extremal dependence, but we
will rather discuss some pitfalls and problems that may arise in
applications in more detail. For simplicity, we mainly consider
bivariate vectors $(X_1,X_2)$ of claim sizes (or risks).

Analogously to our basic assumption \eqref{eq:excessconv} in the
univariate setting, in classical multivariate extreme value theory
it is assumed that the conditional distribution of the suitably
standardized random vector given that its norm exceeds a high
threshold converges to a non-degenerate limit as the threshold
increases. However,  as the components of the vector need not be of
comparable size, the marginal distributions are usually first
standardized, e.g., to the standard Pareto distribution:
\begin{equation} \label{eq:marginnorm}
  Y := \Big(\frac 1{1-F_l(X_l)}\Big)_{1\le l\le 2}.
\end{equation}
Here we assume for simplicity that the marginal cdf's $F_l$, $1\le
l\le 2$, of $X$ are continuous.

Now fix some norm $\|\cdot\|$ on $\R^2$ and suppose that
\begin{equation}  \label{eq:multiregvar}
  P\Big( \frac Yu\in\cdot \,\Big|\, \|Y\|>u\Big) \;\longrightarrow\;
  P^Z(\cdot) \quad \text{weakly}
\end{equation}
as $u\to\infty$ for some non-degenerate limit distribution $P^Z$. In
that case, $Y$ and $P^Y$ are said to be {\em multivariate regularly
varying}.

It can be shown that this condition does not depend on the specific
choice of the norm, while the exact form of the limit distribution
does. For example, if one works with the maximum norm
$\|\cdot\|_\infty$, then \eqref{eq:multiregvar} is equivalent to
\begin{eqnarray} \label{eq:multiregvarmaxnorm}
 P(Y_1>uy_1 \text{ or } Y_2>uy_2 \mid Y_1>u \text{ or } Y_2>u) & = &
\frac{1-F_Y(uy_1,uy_2)}{1-F_Y(u,u)} \nonumber\\
 & \to &
P\{Z_1>y_1 \text{ or } Z_2>y_2\} \nonumber\\
 & =: & 1-H(y_1,y_2)
\end{eqnarray}
for all points $(y_1,y_2)\in [1,\infty)^2$  of continuity of $H$.

Applying the (generalized) polar transformation
$$ \Xi : y\mapsto\Big(\|y\|,\frac y{\|y\|}\Big),
$$
one can conclude from  \eqref{eq:multiregvar} that
$$ P\Big( \Big(\|Y\|,\frac Y{\|Y\|}\Big) \in\cdot \,\Big|\, \|Y\|>u\Big) \;\longrightarrow\;
  P^{\Xi(Z)}(\cdot) \quad \text{weakly.}
$$
Now standard arguments from the theory of regular variation show
that the limiting distribution $P^{\Xi(Z)}$ must be a product
measure with first factor equal to the standard Pareto distribution
and the second factor being some distribution $\Phi$ on the upper
right quadrant $S^+:=\{z\in [0,\infty)\mid \|z\|=1\}$ of the unit
``circle'' with respect to the norm $\|\cdot\|$. Unlike in the
univariate setting, all possible limit distributions do not form a
parametric family, because the so-called {\em spectral probability
measure} $\Phi$ can be any distribution on $S^+$ satisfying the
condition
\begin{equation}  \label{eq:spectralcond}
\int_{S^+} \omega_1\,\Phi(d\omega)=\int_{S^+} \omega_2\,
\Phi(d\omega),
\end{equation}
that follows from the fact that all marginal cdf's are standard
Pareto.
\begin{remark}
  In the literature, instead of \eqref{eq:multiregvar}
  often  the equivalent assumption
  \begin{equation}  \label{eq:multiregvar2}
    u P\Big\{\frac Yu\in\cdot\,\Big\} \,\longrightarrow\, \nu \quad
    \text{vaguely as } u\to\infty
  \end{equation}
  is considered, where $\nu$ is some Radon measure on
  $[0,\infty]^2\setminus\{(0,0)\}$ (see e.g.\ Resnick (2007), Section
  6.1). Similarly as before, one can conclude that the measure
  $\nu^\Xi$ induced by the polar transformation is the product of the measure
  with Lebesgue density $t\mapsto t^{-2}$ on $(0,\infty)$ and a
  finite measure $\tilde\Phi$ on $S^+$, the so-called {\em spectral
  measure}. The latter is related to the spectral probability
  measure via
  $$ \Phi=\frac{\tilde\Phi}{\tilde\Phi(S^+)} \quad \text{and} \quad
  \tilde\Phi = \frac\Phi{\int_{S^+} \omega_1\,\Phi(d\omega)}.
  $$
\end{remark}

For an arbitrary measurable set $A\subset (0,\infty)^2$ one has
\begin{equation}  \label{eq:probabconv}
  P\Big(\frac Yu\in A\,\Big |\, \|Y\|>u\Big) \;\longrightarrow\;
  \int_{S^+}\int_1^\infty 1_{A}(t\omega)
  t^{-2}\, dt\, \Phi(d\omega)
\end{equation}
as $u\to\infty$, provided the set $A$ is continuous with respect to
the limit measure (that is the right-hand side equals 0 if $A$ is
replaced with its topological boundary). Hence, to estimate the
probability that some future observation $X$ falls into a given
extreme set $C$ (e.g., $C=(x_1,\infty)\times (x_2,\infty)$
describing the event that the claim sizes in both lines of business
exceed a given high threshold) can be estimated as follows:
\begin{enumerate}
 \item Estimate the marginal cdf's $F_l$ by $\hat F_l$; for the
 estimation of the tails, the  methods discussed in the previous
 sections can be used.
 \item Transform the data $X_i=(X_{i,1},X_{i,2})$, $1\le i\le n$,
 and the extreme set $C$ using the fitted marginal cdf's:
 \begin{eqnarray*}
   \hat Y_i & := & \Big(\frac 1{1-\hat F_1(X_{i,1})}, \frac 1{1-\hat
 F_2(X_{i,2})}\Big)\\
   \frac 1{1-\hat F(C)} & := & \Big\{ \Big(\frac 1{1-\hat F_1(x_1)}, \frac 1{1-\hat
 F_2(x_2)}\Big)\,\Big |\, (x_1,x_2)\in C\Big\}.
 \end{eqnarray*}
 \item Fix some norm and estimate the corresponding spectral
 probability measure $\Phi$ by $\hat\Phi$ (see below).
 \item Use \eqref{eq:probabconv} and  the  regular variation of
 $\|Y\|$ to approximate
 \begin{eqnarray*}
   P\{X\in C\} & = & P\Big\{ Y\in \frac 1{1-F(C)}
   \Big\}\\
   & = & P\Big( Y\in \frac 1{1- F(C)}\,\Big|\,
   \|Y\|>ru\Big)\cdot \frac{P\{\|Y\|>ru\}}{P\{\|Y\|>u\}} \cdot
   P\{\|Y\|>u\}\\
   & \approx & \int_{S^+}  \int_1^\infty 1_{ 1/(1-F(C))}(t\omega)
  t^{-2}\, dt\, \Phi(d\omega)\cdot r^{-1} \cdot P\{\|Y\|>u\}\\
  & \approx & \int_{S^+}  \int_1^\infty 1_{1/(1-\hat F(C))}(t\omega)
  t^{-2}\, dt\, \hat\Phi(d\omega)\cdot r^{-1}\cdot \frac
  1n\sum_{i=1}^n 1_{\textstyle\{\|\hat Y_i\|>u\}}.
 \end{eqnarray*}
 Here we have assumed that $\|y\|>ru$ for all $y\in 1/(1-F(C))$, while on the other hand
 $ru$ is sufficiently large such that
 \eqref{eq:probabconv} (with $ru$ instead of $u$) yields a good
 approximation. Moreover,  on the one hand $u$ must be sufficiently large
 such that $P\{\|Y\|>ru\}/P\{\|Y\|>u\}\approx r^{-1}$, while on the
 other hand it must be sufficiently small such that $P\{\|Y\|>u\}$
 can be well estimated by its empirical counterpart.
\end{enumerate}

As the family of all possible spectral (probability) measures is
nonparametric (infinite dimensional), it is substantially more
complicated to estimate $\Phi$ (or $\tilde\Phi$) than to fit the
tail of a univariate cdf. In the last decade, though, a variety of
nonparametric estimators of the spectral measure and related
functions, that also characterize the extremal dependence structure,
have been suggested and analyzed; see, for instance, de Haan and
Ferreira (2006), Chapter 7, Beirlant et al.\ (2004), Chapter 9, or
Einmahl and Segers (2009). These estimators also use marginally
transformed observations $\hat Y_i$, but, unlike in program to
estimate the probability of extreme events outlined above, here one
may use fully nonparametric estimators of the marginal cdf's, which
amounts to working with the coordinatewise ranks of the original
observations. In the analysis of the asymptotic behavior of these
estimators, it is important to {\em not} assume that the marginal
cdf's are known, because usually the transformation of the data
using {\em estimated} marginal cdf's (rather than the true ones)
does have a non-negligible influence on the estimation error for
$\Phi$; ignoring it may lead to a wrong assessment of the estimation
accuracy (see, e.g., Einmahl and Segers (2009), p.\ 2960).

As an alternative approach, it has been suggested to  assume some
parametric submodel of spectral (probability) measures and then to
fit this model to the transformed observations using maximum
likelihood or a generalized moment method. Sometimes it is even
assumed that, for some $\Phi$ from this parametric family,  in
\eqref{eq:probabconv} equality holds for some sufficiently large
$u$. We consider this approach quite problematic, because it will
rarely be possible to argue for a specific parametric family of
spectral measures based on either experience with similar data sets
or some ``physical'' reasoning about the process generating the
extremal dependence. Instead, the parametric families are almost
always chosen with mathematical convenience in view, while it is
argued that the family is sufficiently flexible to capture many
different dependence structures. Then, however, one merely trades
the random estimation error, that can be quantified in the
nonparametric framework, for the risk of a model misspecification,
that can hardly be assessed. Hence the seemingly increased
estimation accuracy which comes with the parametric approach {\em if
the model is correct} will often be just a chimera, which possibly
leads to an assessment of the insured risk which is not prudent
anymore.

An even more restrictive modeling approach is related to a
reformulation of the multivariate regular variation in terms of
copulas. Any multivariate cdf $F$ with marginal cdf's $F_l$, $1\le
l\le d$, can be represented as $F(x_1,\ldots,x_l)=C(F_1(x_1),\ldots,
F_d(x_d))$ where $C$ is  a so-called copula, i.e.\ a multivariate
cdf with uniform margins. If all $F_l$ are continuous, then $C$ is
unique: it is the joint cdf of $(F_1(X_1),\ldots, F_d(X_d))$. Thus,
the cdf $F_Y$ of $Y$ defined by \eqref{eq:marginnorm} equals
$F_Y(y_1,y_2)=C(1-1/y_1,1-1/y_2)$ and convergence
\eqref{eq:multiregvarmaxnorm} is equivalent to
$$ \lim_{t\downarrow 0}
\frac{1-C(1-ty_1,1-ty_2)}{1-C(1-t,1-t)}=1-H(y_1^{-1},y_2^{-1})
$$
for all points $(y_1^{-1},y_2^{-1})$ of continuity of $H$. Hence the
parametric approach outlined above boils down to assuming that, on a
small neighborhood of the point (1,1), the copula $C$ can be well
approximated by a  parametric model.

Because the estimation of a general copula is essentially as
difficult as the estimation of a general multivariate cdf (in
particular it is also plagued by the well-known ``curse of
dimensionality''), it has been suggested to assume parametric models
for the whole copula $C$. This approach, however, does not only
introduce a modeling error that is difficult to quantify, but it
contradicts the general philosophy of extreme value theory to ``let
the tails speak for themselves''. In contrast, while the estimation
error of the aforementioned nonparametric estimators of the extremal
dependence structure can be quite large (in particular in higher
dimensions), it can be well assessed under weak model assumptions
and thus it can be taken into account by the risk manager. For that
reason, with the rare exception of those situations when there are
convincing arguments that a particular family of copulas contains
the true one (and not just a crude approximation to it), one should
take the utmost care in analyzing the tail risk using parametric
copulas, in particular in actuarial applications where prudence is a
time-honored principle. (In this context, the interested reader is
advised to study the article by Mikosch (2006) and the pertaining
discussion for an enlightening and entertaining argument about the
pros and cons of copula modeling.)

While the program sketched above in four steps will often yield a
reasonable assessment of the risk that a future observation falls
into some given extreme set if some nonparametric estimator of
$\Phi$ is used, there is one important case in which the result can
be quite misleading. Suppose that large values of one component of
the transformed vector $Y$ of risks do not usually coincide with
large values of the other component, or more precisely that
\begin{equation}  \label{eq:asindep}
   P(Y_2>u\mid Y_1>u) \;\to\; 0
\end{equation}
as $u\to\infty$. In that case, $X_1$ and $X_2$ (or $Y_1$ and $Y_2$)
are said to be {\em asymptotically independent}. Straightforward
calculations show that then the limit measure $\nu$ in
\eqref{eq:multiregvar2} has no mass on $(0,\infty]^2$, i.e., it is
concentrated on the axes. Hence, $\tilde\Phi$ and $\Phi$ have mass
only in the points $(0,1)$ and $(1,0)$ if one uses one of the usual
$p$-norms on $\R^2$ with $p\in[1,\infty]$. (Indeed, because of the
normalization constraint \eqref{eq:spectralcond}, $\Phi$ must be the
uniform distribution on $\{(0,1),(1,0)\}$.) But then for all sets
$A$ that do not intersect with any of the axes, the limit in
\eqref{eq:probabconv} is 0, which usually is too crude an
approximation for the left-hand side.

Note that asymptotic independence is a property of the copula of
$X$. Many popular parametric families of copulas allow for
asymptotic independence for suitable parameter values. A thorough
analysis of the tail behavior of so-called Archimedean copulas both
in the case of asymptotic independence and of asymptotic dependence
can be found in Charpentier and Segers (2009).

To obtain more useful approximations, one has to specify the rate at
which $P(Y_2>u\mid Y_1>u)$ tends to 0. More precisely, one assumes
that for some nontrivial function $d$
\begin{equation} \label{eq:LedfordTawn}
  \frac{P\{Y_1>uy_1, Y_2>uy_2\}}{P\{Y_1>u, Y_2>u\}} \;\to\;
  d(y_1,y_2)
\end{equation}
as $u\to\infty$ uniformly for all points $(y_1,y_2)$ with
$\max(y_1,y_2)=1$. As an immediate consequence, one obtains the
regular variation of the function $u\mapsto
P\{Y_1>u,Y_2>u\}=P\{\min(Y_1,Y_2)>u\}$:
\begin{equation}  \label{eq:etadef}
 \frac{P\{\min(Y_1,Y_2)>ux\}}{P\{\min(Y_1,Y_2)>u\}} \;\to\; x^{-1/\eta}
\end{equation}
as $u\to\infty$ for all $y_1,y_2>0$ and some $\eta\in(0,1]$ which is
called the {\em coefficient of tail dependence}. Moreover, the
limiting function $d$ is homogeneous of order $-1/\eta$. Since $Y_1$
and $Y_2$ are asymptotically independent whenever
$P\{Y_1>u,Y_2>u\}=o(u)$, in particular one has asymptotic
independence if $\eta<1$. If $Y_1$ and $Y_2$ are {\em exactly}
independent, then $\eta=1/2$, while, roughly speaking, values of
$\eta$ between 1/2 and 1 indicate a positive, but asymptotically
vanishing dependence between the large values of $Y_1$ and $Y_2$. A
slight modification of this model was first suggested by Ledford and
Tawn (1996,1997).

To construct estimators of the coefficient of tail dependence, let
$\hat Y_{i,l}^{(n)} := (n+1)/(n+1-R_{i,l})$, $1\le i\le n, 1\le l\le
2$, with $R_{i,l}$ denoting the rank of $X_{i,l}$ among
$X_{1,l},\ldots,X_{n,l}$. Hence $\hat Y_{i,l}^{(n)}=1/(1-\hat
F_l(X_{i,l}))$ where $\hat F_l$ is essentially the empirical cdf of
$X_{1,l},\ldots,X_{n,l}$, with a minor modification to avoid
division by 0. In view of \eqref{eq:etadef}, the rv's
\begin{equation} \label{eq:Tindef}
  T_i^{(n)} := \min\big(\hat Y_{i,1}^{(n)},\hat Y_{i,2}^{(n)}\big)
\end{equation}
have approximately a Pareto tail with extreme value index $\eta$,
that can be estimated by one of the usual estimators discussed in
Section \ref{sect:tailanalysis} applied to $T_i^{(n)}$, $1\le i\le
n$. Draisma et al.\ (2004) proved an analog to Theorem
\ref{theo:Qnlimit} for the tail empirical quantile function
pertaining to these rv's. It turned out that in case of asymptotic
independence one obtains the same limit as in Theorem
\ref{theo:Qnlimit}, so that also all the results on the extreme
value estimators discussed in Section \ref{sect:tailanalysis} carry
over to the present situation (although the rv's $T_i^{(n)}$ are not
exactly independent). If the components are not asymptotically
independent, one can still conclude the asymptotic normality of the
estimators of $\eta$, but the formulas for the asymptotic variance
are more complicated and depend on the positive limit of
$P(Y_2>u\mid Y_1>u)$.

Drees and M\"{u}ller (2008) proposed the estimator
\begin{equation}  \label{eq:dest}
 \hat d(y_1,y_2) := \frac 1m \sum_{i=1}^n 1_{\textstyle\{\hat
Y_{i,1}>T^{(n)}_{n-m:n}y_1, \hat Y_{i,2}>T^{(n)}_{n-m:n}y_2\}}
\end{equation}
 of
the limiting function $d(y_1,y_2)$ in \eqref{eq:LedfordTawn} and
proved uniform convergence of the suitably standardized estimation
error $\hat d-d$ towards a centered Gaussian process under suitable
smoothness conditions on $d$. Moreover, they derived statistical
tests and a graphical tool to validate the model assumption
\eqref{eq:LedfordTawn}. Finally, the theory was applied to two
well-known bivariate data sets of claim sizes, the first describing
losses to buildings and losses to their content in Danish fire
insurances, while the second was taken from the Society of Actuaries
Group Medical Insurance Large Claims Database (cf.\ Section
\ref{sect:dataanalysis}). In both cases it seems very likely that
condition \eqref{eq:LedfordTawn} holds with a coefficient of tail
dependence $\eta$ less than 1. As the point estimates of $\eta$ were
larger than 1/2, the claim sizes in the different lines of business
are probably asymptotically independent, though with a
non-negligible positive dependence for finite levels. Applying
classical bivariate extreme value theory in such cases will usually
lead to a wrong assessment of the true risk insured.

So far, the extremal dependence structure has been analyzed in terms
of the joint distribution of the observations after a
standardization of the marginals to some fixed distribution
(standard Pareto or uniform). Consequently both coordinates of the
random vector have been treated symmetrically. As an alternative, in
recent years the so-called conditional extreme value (CEV) approach
has been considered, where the asymptotic behavior of one component
(after a suitable linear normalization) given that the other
component is large is investigated. For example, Abdous et al.\
(2005) and Abdous et al.\ (2008) considered the limit behavior of
\begin{equation} \label{eq:condprob}
  P\big( X_2\le a(x_i)x_2+b(x_1)\mid X_1>x_1\big)
\end{equation}
as $x_1\to\infty$ for elliptically distributed  vectors $(X_1,X_2)$,
and Foug\`{e}res and Soulier (2010) determined such limit distributions
for more general bivariate distributions in polar representation.
Das and Resnick (2011) examined possible limits in
\eqref{eq:condprob} in a general framework of regular variation on
cones and discussed the relationship to the approach by Ledford and
Tawn. In the special case $a(x_1)=1$ and $b(x_1)=0$ (coined standard
regular varying case by Das and Resnick), the CEV approach
facilitates the analysis of the extremal dependence between large
claim sizes in absolute terms rather than just a comparison the
behavior of the conditional distribution of one claim size given
that the other is large relative to its unconditional distribution
as in the approach by Ledford and Tawn. Unfortunately, in most
applications one needs different normalizing functions $a$ and $b$
to obtain a non-degenerate limit of \eqref{eq:condprob}.

It is worth mentioning that the methods outlined in this section are
only useful to analyze the dependence between extreme claim sizes
(or losses) observed in a small number of different lines of
insurances (usually sold to the same customer). From an economic
point of view, the dependence between different risks insured in the
same line of business is often more important. For example, in
property insurance a large storm will usually result in many claims
from customers living in the same area. With the present state of
the art, extreme value theory has little to offer to analyze the
extremal dependence in such situations. Instead, approaches using
expert knowledge (e.g., from meteorology) remain the methods of
choice.

\section{Analyzing Large Claim Sizes in Health Insurance}
  \label{sect:dataanalysis}

In the years 1991 and 1992 the Society of Actuaries collected large
claim sizes (totalling \$25,000 or more)  in US health insurances.
The resulting large claim size database is available at the website
\textit{http://www.soa.org}. For each claimant, hospital charges and
other charges were recorded in each year together with the type of
the health insurance plan and the status of the claimant (employee
or dependent), among other information. See Grazier and G'Sell
Associates (1997) for a detailed description of the data set (see
also Cebri\'{a}n et al.\ (2003) for a statistical analysis of the total
charges).

A closer inspection of the data reveals that the structure of the
claim sizes depend on the status of the claimant and the type of the
health insurance plan. For example, the large non-hospital costs
were significantly larger for HMO (health maintenance organization),
POS (point of service) and indemnity plans than for PPO (preferred
provider organization), EPO (exclusive provider organization),
comprehensive and other indemnity plans as well as for those records
for which the type of plan was unknown. Therefore, as an example
here we analyze the claims for the second group of health plans  in
the year 1992 when the claimant had the status `dependent'.

The sampling scheme that only those claims with total costs of at
least \$25,000 were recorded introduces an artificial negative
dependence between both components: if the hospital charges were
small, say less than \$5,000, then the other charges must be larger
than \$20,000, and vice versa. For that reason, here we only
consider those records for which both type of charges were at least
\$25,000, leading to a sample of size $n=1959$. (We discuss the
consequences of this choice at the end of the section.)

\begin{figure}[htbp]
\includegraphics[width=180mm]{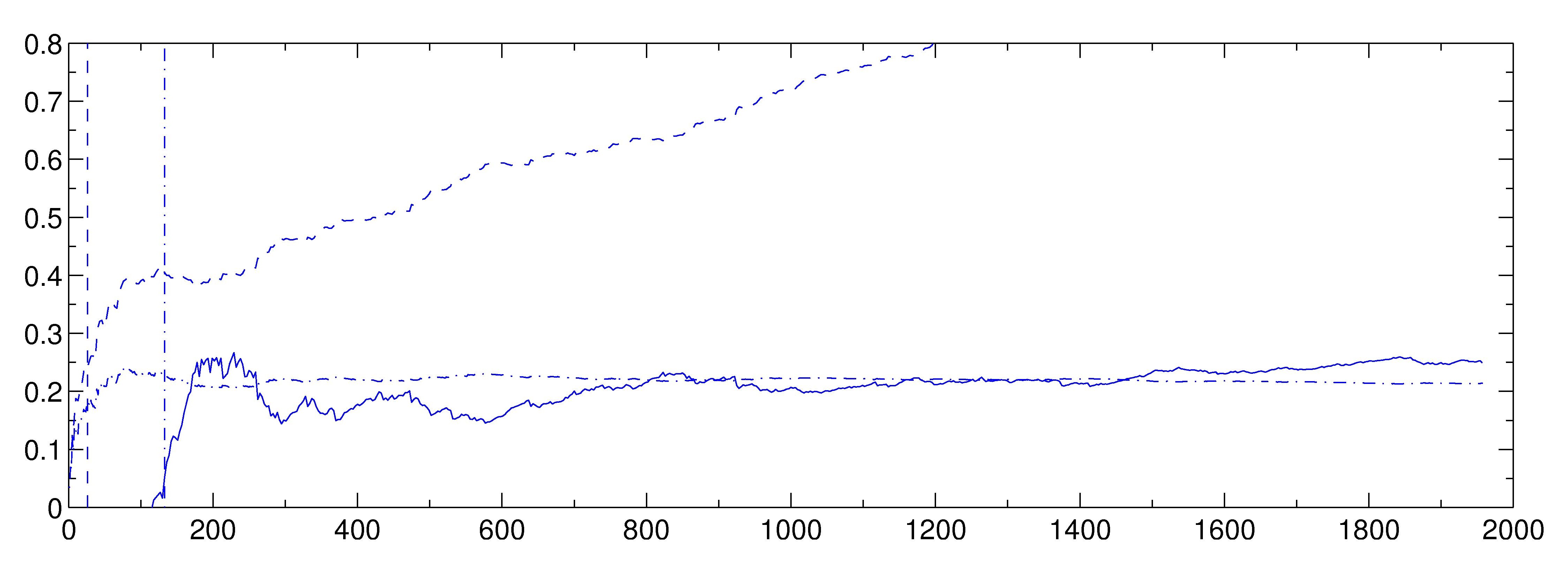}


 \caption{ML estimator (solid line) and Hill
estimator (dashed line) and the Hill estimator applied to the
 data shifted by \$ 300,000 (dash-dotted line) based on $k+1$ largest hospital charges versus
 $k$; the estimated optimal number $\hat k_n^{DK}$ for the Hill
 estimators are indicated by vertical lines}
 \label{fig:gammahosp}
\end{figure}

First we fit Pareto distributions to the marginal tails. Figure
\ref{fig:gammahosp} shows the maximum likelihood estimator (in the
GPD model) for the extreme value index of the hospital charges
(solid line) and the Hill estimator (dashed line) as a function of
$k$ (i.e., the number of largest observations used for estimation
reduced by 1). Obviously, the graph of the ML estimator is much more
stable than that of the Hill estimator. While for the former it
seems reasonable to use at least $1400$ largest observations, the
Hill plot increases more or less monotonically for $k\ge 300$, say.
Moreover, the data-driven procedures to select an optimal number of
order statistics yield very small values of $k$ for the Hill
estimator ($\hat k_n^{boot}=38$, $\hat k_n^{DK}=26$).

 As explained in Example
\ref{ex:shift}, this qualitatively different behavior may be due to
a constant term in the GPD approximation of the tail, that does not
influence the shift invariant ML estimator but that leads to a large
bias of the Hill estimator if $k$ is chosen too large. In fact, the
ML estimator fits a shifted Pareto distribution with location
parameter of about $-3\cdot 10^5$ for a wide range of $k$-values. If
one adds \$300,000 to each of the hospital charges, then the Hill
plot (displayed by the dash-dotted line in Figure
\ref{fig:gammahosp}) becomes very flat so that at first glance the
Hill plot suggests that one may use almost all observations to
estimate $\gamma$. So apparently for this data set the instability
of the Hill estimator is indeed largely due to its sensitivity to
shifts. For the shifted data the optimal number of order statistics
estimated by the bootstrap and the sequential procedure sketched in
Section \ref{sect:validate} suggest to use the 143 and 134 largest
order statistics, respectively, resulting in a Hill estimate of
about 0.22.

\begin{figure}[tbp]
\centerline{\epsfxsize=12cm \epsfbox{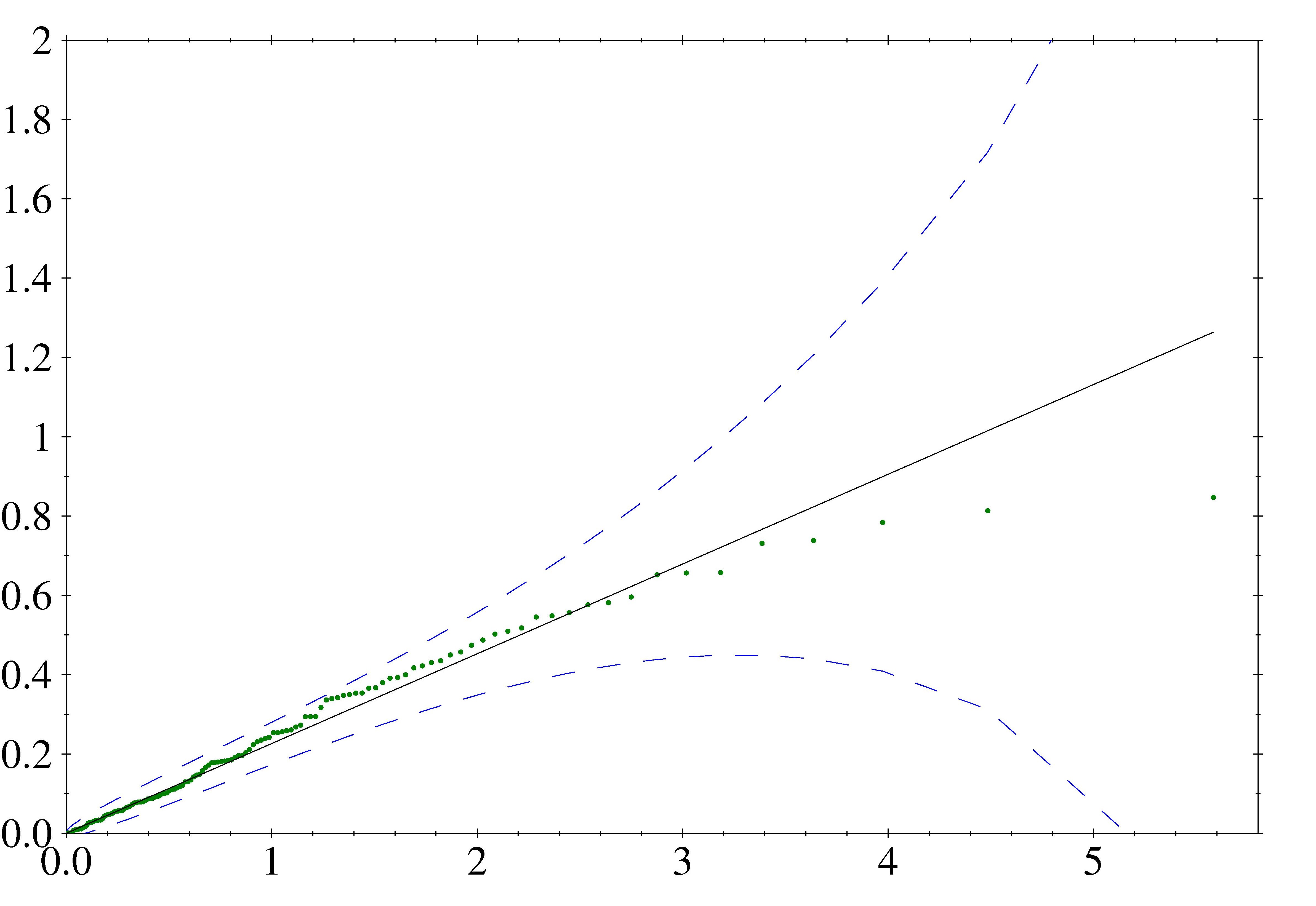}}
\vspace*{-0.4cm} \caption{Pareto-qq-plot of 133 largest hospital
charges together with 0.95\%-``confidence band''}
 \label{fig:qqplot}
\end{figure}

Figure \ref{fig:qqplot} shows the qq-plot for the shifted data
together with the line with slope equal to the Hill estimate
$\hat\gamma_{n,k}^{(1)}\approx 0.286$ for $k=133$. Moreover, the
functions $-\hat\gamma_{n,k}^{(1)}\big(\log t \pm c_\alpha
\big(t^{-1}-1-(\log t)^2\big)^{1/2}(t(1-t))^{-1/10}\big)$ are
displayed as dashed lines. Here, $c_\alpha=2.78$ was calculated by
Monte Carlo simulations as described in Section \ref{sect:validate}
such that the probability that all points of the qq-plot lie between
these graphs is about $1-\alpha=0.95$. (Strictly speaking, the
probability is probably a bit higher, because the band does not take
into account the fact that the shift has been chosen depending on
the data to improve on the Pareto fit.) Indeed, the fit is
reasonably good and all points lie in the band bordered by these
functions. However, for the most extreme points the qq-plot flattens
out, indicating that perhaps the fitted Pareto tail slightly
overstates the actual risk, which may be desirable for a prudent
risk assessment. (If one uses the largest 1900 observations as
suggested by a superficial inspection of the Hill plot, then the
Hill estimate is not changed much and still all points of the
qq-plot lie within the confidence band.)

\begin{figure}[htbp]
\centerline{\epsfxsize=12cm \epsfbox{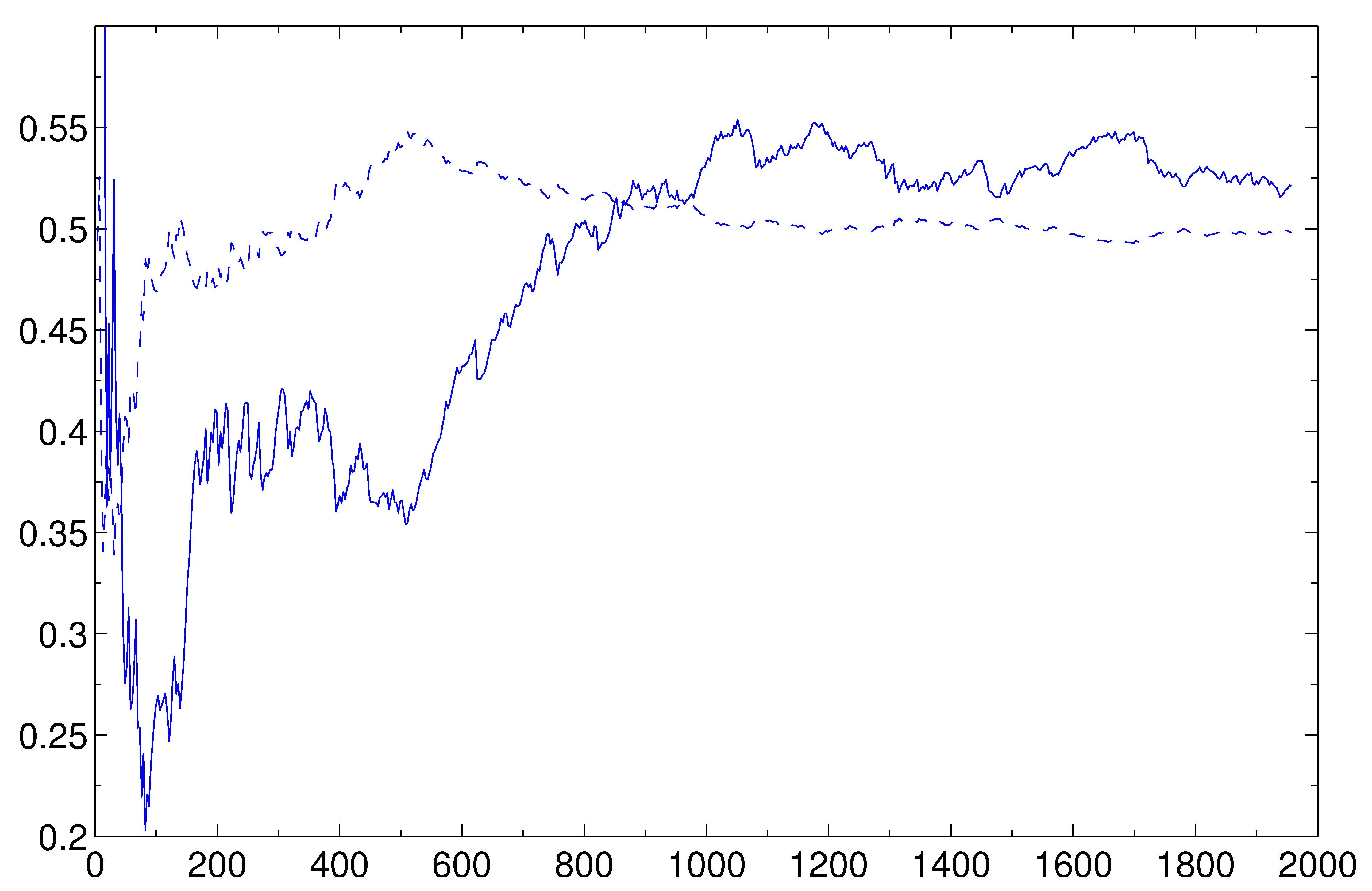}} \caption{ML
estimator (solid line) and Hill estimator (dashed line) based on
$k+1$ largest other charges versus $k$}
 \label{fig:gammaother}
\end{figure}

Figure \ref{fig:gammaother} displays the ML estimator and the Hill
estimator for the second component describing the other costs. These
plots are less stable than the ones for the hospital charges. For
both estimators it seems certainly advisable to not use many more
than 350 largest observations to fit the tail; the bootstrap and the
sequential procedure suggest to merely use $\hat k_n^{boot}=125$ and
$\hat k_n^{DK}=90$ largest order statistics for the Hill estimator.
As Hill estimate one obtains $\hat\gamma_{n,125}^{(2)}\approx 0.495$
and a $95\%$-confidence interval of about $[0.41,0.58]$. Hence,
apparently the other charges are significantly heavier tailed than
the hospital charges.

\begin{figure}[htbp]
\centerline{\epsfxsize=10cm \epsfbox{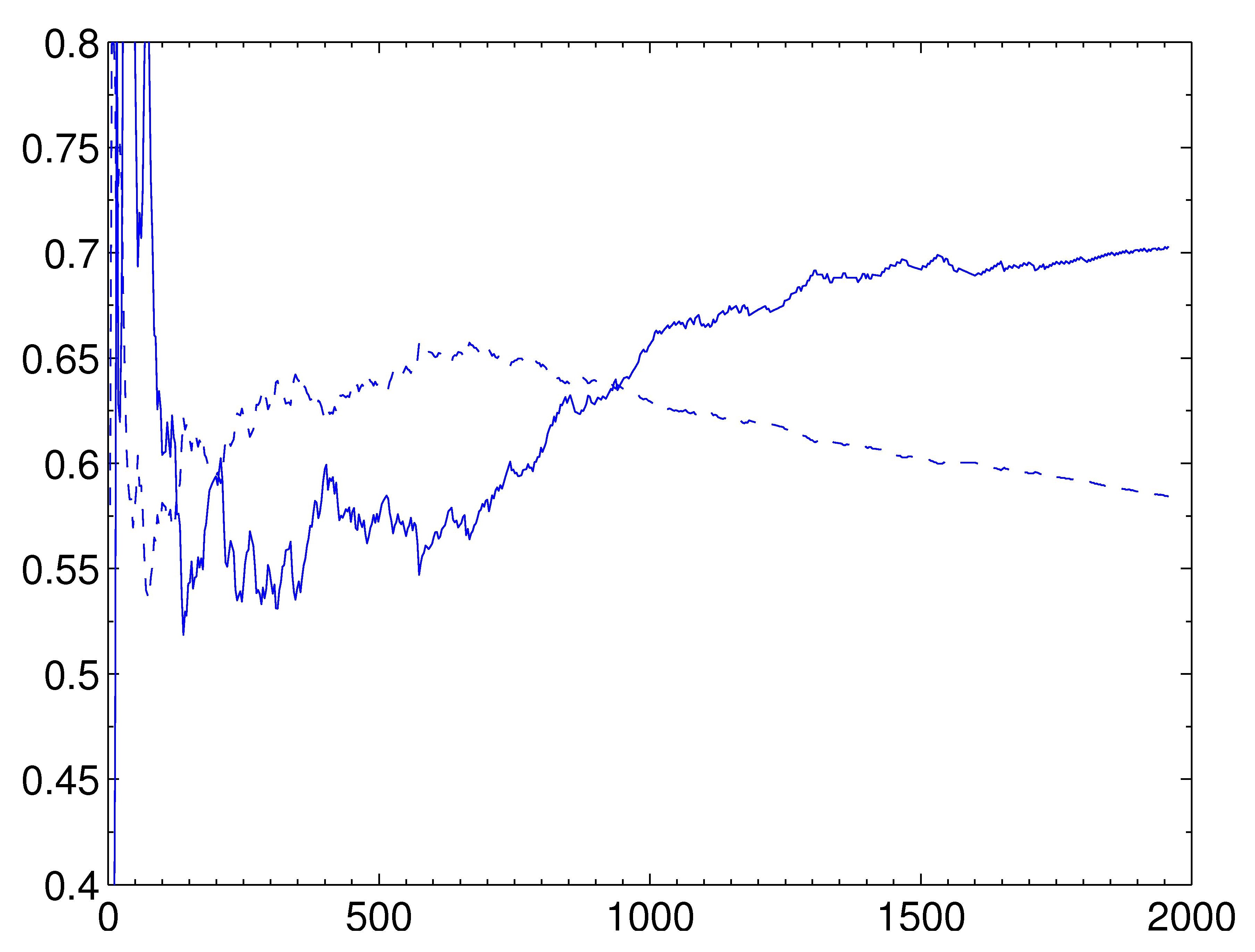}} \caption{ML
estimator (solid line) and Hill estimator (dashed line) for $\eta$
based on $k+1$ order statistics of $T_i^{(n)}$ versus $k$}
 \label{fig:etaplot}
\end{figure}

The ML estimator and the Hill estimator of the coefficient $\eta$ of
tail dependence between both types of costs (based on the rv's
$T_i^{(n)}$ defined by \eqref{eq:Tindef}) are shown in Figure
\ref{fig:etaplot}. Here perhaps up to 600 large order statistics of
the $T_i^{(n)}$ can be used for the ML estimator. Note that the
mathematical theory for the data-driven procedures of choosing $k$
has only been developed for the Hill estimator based on iid data.
Hence, strictly speaking, it is not applicable here, but the
aforementioned result by Draisma et al.\ (2004) suggests that in the
present situation the sequential estimator, that yields $\hat
k_n^{DK}=318$, has the same asymptotic behavior as for iid data. The
resulting Hill estimate 0.63 hints at a rather weak, asymptotically
vanishing, but non-negligible dependence, because the pertaining
confidence interval [0.52, 0.74] does neither contain 1 or 1/2.

\begin{figure}[htbp]
\centerline{\epsfxsize=14cm \epsfbox{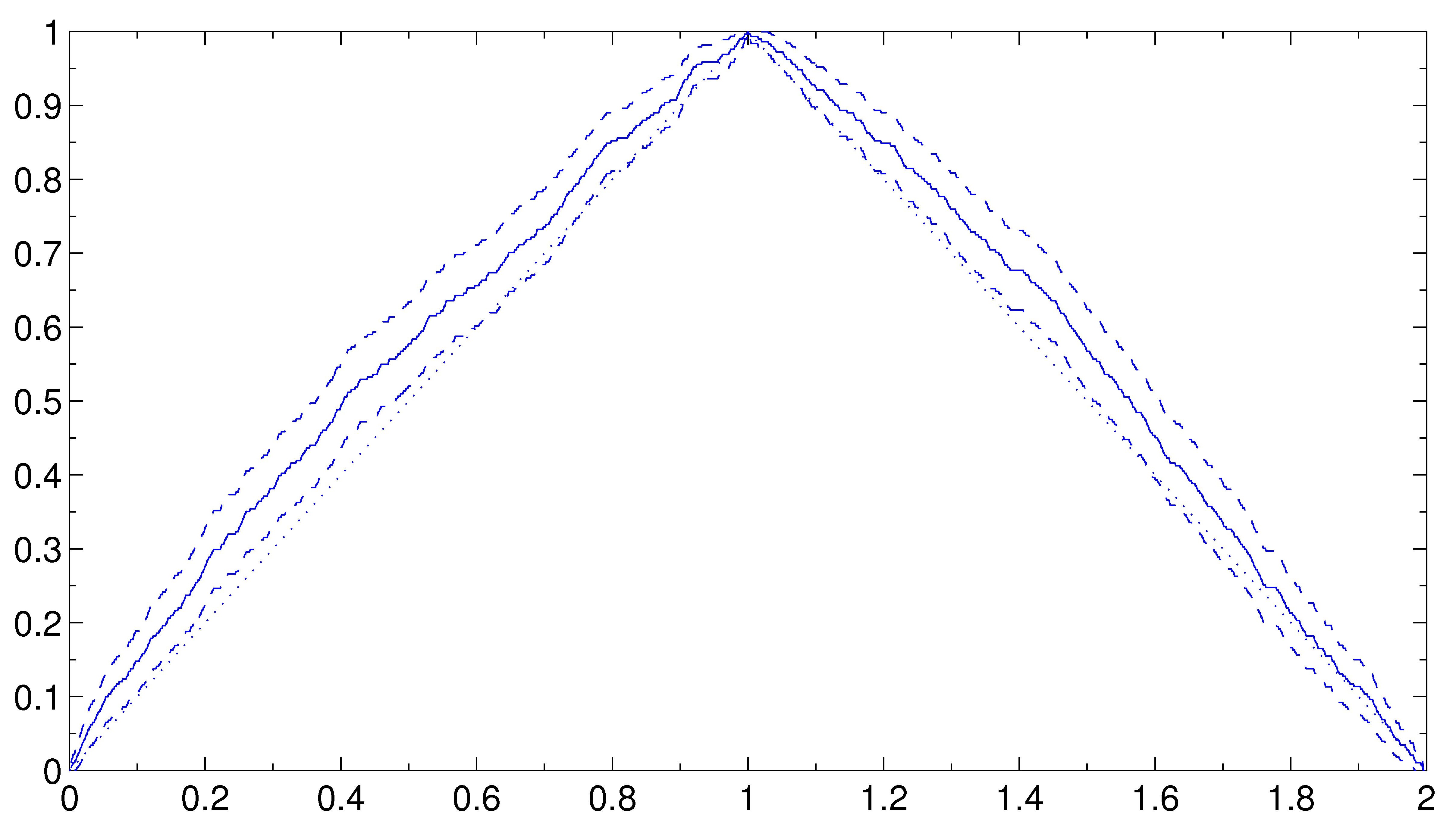}}
\caption{Estimator of $d(1/x,1)$ ($0<x\le 1$) and $d(1,1/(2-x))$
($1\le x<2$)(solid line); pointwise asymptotic $95\%$-confidence
intervals are indicated by dashed lines; for comparison, the lines
$x\mapsto x$ and $x\mapsto 2-x$ are shown by dotted lines}
 \label{fig:dest}
\end{figure}

Finally, we consider the estimate $\hat d(y_1,y_2)$ defined in
\eqref{eq:dest}. Figure \ref{fig:dest} depicts the estimates
$$ x\mapsto \left\{ \begin{array}{l@{\quad}l}
  \hat d(1/x,1), & 0<x\le 1,\\
  \hat d(1,1/(2-x)), & 1\le x<2.
  \end{array} \right.
$$
From these values and the estimate $\hat\eta_n$ one can calculate
estimates of $d(y_1,y_2)$ for all values $y_1,y_2>0$ because of the
homogeneity of $d$ of order $-1/\eta$:
$$d(y_1,y_2)=(\min(y_1,y_2))^{-1/\eta}
d\Big(\frac{y_1}{\min(y_1,y_2)},\frac{y_2}{\min(y_1,y_2)}\Big).
$$

In view of \eqref{eq:LedfordTawn}, one may approximate
\begin{eqnarray*}
 d\Big(\frac 1x,1\Big) & \approx & P\Big( Y_{i,1}>\frac ux\,\Big|\,
Y_{i,1}>u,Y_{i,2}>u\Big)\\
& = & P\Big( X_{i,1}>F_1^\leftarrow\Big(1-\frac xu\Big)\,\Big|\,
X_{i,1}>F_1^\leftarrow\Big(1-\frac
1u\Big),X_{i,2}>F_2^\leftarrow\Big(1-\frac 1u\Big)\Big)
\end{eqnarray*}
for large $u$. Observe that in Figure \ref{fig:dest}, the estimate
of this probability is just slightly larger than
$x=P(Y_{i,1}>u/x\mid Y_{i,1}>u)=P(X_{i,1}>F_1^\leftarrow(1-x/u)\mid
X_{i,1}>F_1^\leftarrow(1-1/u))$. Hence, given that the hospital
charge exceeds a high threshold, the fact that also the other
charges exceed an analogously high threshold does not alter the
conditional distribution of the hospital charges much. The same
observation can be made with the roles of hospital charges and other
charges interchanged. This property should not be confused with the
asymptotic independence condition \eqref{eq:asindep} in which one
conditions only at the event that one component is large. Indeed, it
can be shown that to each $\eta\in (0,1)$ and each function
$g:[0,2]\to[0,1]$ that is increasing on $[0,1]$ and decreasing on
$[1,2]$ with $g(0)=g(2)=0$ and $g(1)=1$, one can find a probability
distribution $P^Y$ such that \eqref{eq:LedfordTawn} holds with
$$ d(y_1,y_2) = \left\{ \begin{array}{l@{\quad}l}
y_2^{-1/\eta} g(y_1/y_2,1), & y_1\ge y_2,\\
y_1^{-1/\eta} g(1,y_2/y_1), & y_1< y_2.
\end{array} \right.
$$
Hence the function whose estimate is shown in  Figure \ref{fig:dest}
can be combined with any value of the coefficient of tail dependence
in $(0,1)$ to obtain a limiting function $d$ in
\eqref{eq:LedfordTawn}. (The converse result that the function $d$
can be represented in such a way is an easy consequence of its
homogeneity; cf.\ e.g.\ Charpentier and Juri, 2006, Remark 3.4.)

As we have considered only those claims for which both components
are at least \$25,000, in fact we have analyzed the conditional
distribution of the claim sizes given that both components are at
least \$25,000. If instead we use all record for which at least one
of the component exceeds \$25,000, then a different conditional
distribution is analyzed. Indeed, for this larger data set one
obtains higher estimates of the coefficient of tail dependence,
indicating a stronger extremal dependence between both types of
charges. At first glance, this fact seems counterintuitive, because
in assumption \eqref{eq:etadef} only probabilities of events occur
in which both components are large. Notice, however, that the
$Y_{i,l}$ have been calculated by transforming $X_{i,l}$ using the
marginal cdf $F_l$, and this marginal distribution is different for
the two conditional settings described above. For that reason, in
contrast to the first impression, the parameter $\eta$ and likewise
the limiting function $d$ also depend on the stochastic behavior of
the vector on the regions where just one of the components is large.

\section{Proofs}
 \label{sect:proofs}

\begin{proofof}  Corollary \ref{cor:premiumestAN}. \rm
 For the sake of simplicity, we assume that $F$ is continuous on
 $(\qF(1-\eta),\infty)$ for some $\eta>0$, but a slight refinement
 of the arguments given below shows that the assertion holds without
 this continuity assumption.
  With
  $$ \psi(t;\gamma) := \int_0^t (1+\gamma x)^{-1/\gamma}\, dx $$
  and $u_n := \qF(1-k_n/n)$, the estimation error can be decomposed
  as follows:
  \begin{eqnarray*}
    \lefteqn{\frac{n/k_n}{\tilde a(k_n/n)} \Big(
    \hat\Pi_n(t_n,c_n)-\int_{t_n}^{t_n+c_n}1-F(s)\, ds\Big)}\\
    & = & \frac{S(Q_n)}{\tilde a(k_n/n)} \Big( \psi\Big(
    \frac{t_n+c_n-Q_n(1)}{S(Q_n)}; T(Q_n)\Big) - \psi\Big(\frac{t_n+c_n-u_n}{\tilde a(k_n/n)};
    T(Q_n)\Big)\Big)\\
    & & - \frac{S(Q_n)}{\tilde a(k_n/n)} \Big( \psi\Big(
    \frac{t_n-Q_n(1)}{S(Q_n)}; T(Q_n)\Big) - \psi\Big(\frac{t_n-u_n}{\tilde a(k_n/n)};
    T(Q_n)\Big)\Big)\\
    & & + \Big( \frac{S(Q_n)}{\tilde a(k_n/n)} -1\Big) \Big( \psi\Big(\frac{t_n+c_n-u_n}{\tilde a(k_n/n)};
    T(Q_n)\Big)- \psi\Big(\frac{t_n-u_n}{\tilde a(k_n/n)};
    T(Q_n)\Big)\Big)\\
    & & + \psi\Big(\frac{t_n+c_n-u_n}{\tilde a(k_n/n)};
    T(Q_n)\Big)- \psi\Big(\frac{t_n-u_n}{\tilde a(k_n/n)};
    T(Q_n)\Big) - \Big(\psi\Big(\frac{t_n+c_n-u_n}{\tilde a(k_n/n)};
    \gamma\Big)- \psi\Big(\frac{t_n-u_n}{\tilde a(k_n/n)};
    \gamma\Big)\Big)\\
    & & + \int_{(t_n-u_n)/\tilde a(k_n/n)}^{(t_n+c_n-u_n)/\tilde
    a(k_n/n)} (1+\gamma x)^{-1/\gamma} - \frac n{k_n}(1-F(u_n+\tilde
    a(k_n/n)x))\, dx\\
    & =: & I_a-I_b+II+III+IV.
  \end{eqnarray*}
  It will turn out that, under the conditions of the corollary, the
  term III dominates all other terms.

  Because
  $$ \frac\partial{\partial \gamma} \psi(t;\gamma) = \int_0^t \frac
  1{\gamma^2} \log(1+\gamma x)(1+\gamma x)^{-1/\gamma}-\frac
  x\gamma(1+\gamma x)^{-(1+1/\gamma)}\, dx
  $$
  (which is interpreted as $\int_0^t x^2\e^{-x}\, dx/2$ for
  $\gamma=0$), the mean value theorem shows that
  $$ III = (T(Q_n)-\gamma) \int_{(t_n-u_n)/\tilde a(k_n/n)}^{(t_n+c_n-u_n)/\tilde
    a(k_n/n)} \frac
  1{\tilde\gamma_n^2} \log(1+\tilde\gamma_n x)(1+\tilde\gamma_n x)^{-1/\tilde\gamma_n}-\frac
  x{\tilde\gamma_n}(1+\tilde\gamma_n x)^{-(1+1/\tilde\gamma_n)}\, dx
  $$
  for some $\tilde\gamma_n$ between $T(Q_n)$ and $\gamma$, which implies $\tilde\gamma_n-\gamma=O_P(k_n^{-1/2})$.

  First consider the case $\gamma>0$. Then $1+\tilde\gamma_n
  x\to\infty$ uniformly over the range of integration, because
  \begin{eqnarray}  \label{eq:tnasymp}
   \frac{y-u_n}{\tilde a(k_n/n)} & = &
  \frac{(n(1-F(y))/k_n)^{-\gamma}-1}\gamma+R\Big(\frac{k_n}n,
  \frac n{k_n}(1-F(y)\Big)\nonumber\\
  & = &  \frac{(n(1-F(y))/k_n)^{-\gamma}-1}\gamma +o\Big(k_n^{-1/2}\tau_n
  \Big(\frac n{k_n}(1-F(y))\Big)^{-\gamma}\Big) \nonumber\\
  & = &  \frac{(n(1-F(y))/k_n)^{-\gamma}-1}\gamma (1+o(1))
  \end{eqnarray}
  uniformly for $y\in[t_n,t_n+c_n]$ by assumptions \eqref{eq:kncond4} and \eqref{eq:kncond3}.
  It follows that $x(1+\tilde\gamma_n x)^{-(1+1/\tilde\gamma_n)}
  =o\big(\log(1+\tilde\gamma_n x)(1+\tilde\gamma_n x)^{-1/\tilde\gamma_n}\big)$ and $\log(1+\tilde\gamma_n
  x)=O(\log(n(1-F(t_n))/k_n))$,  and thus by condition \eqref{eq:kncond3}
  \begin{eqnarray*}
   \lefteqn{\frac{ \log(1+\tilde\gamma_n x)(1+\tilde\gamma_n
  x)^{-1/\tilde\gamma_n}}{  \log(1+\gamma x)(1+\gamma
  x)^{-1/\gamma}}}\\
  & = & \bigg(
  1+\frac{\log\big(1+\frac{(\tilde\gamma_n-\gamma)x}{1+\gamma x}\big)}{\log(1+\gamma
  x)}\bigg)
  \Big(1+\frac{(\tilde\gamma_n-\gamma)x}{1+\gamma x}\Big)^{-1/\tilde\gamma_n}
  \exp\Big(\Big(\frac 1\gamma-\frac
  1{\tilde\gamma_n}\Big)\log(1+ \gamma x)\Big)\\
  &= & 1+O_P\Big(k_n^{-1/2} \log\Big( \frac
  n{k_n}(1-F(t_n))\Big)\Big)\\
  & = & 1+o_P(1)
 \end{eqnarray*}
 uniformly over the range of integration.
 Therefore, in the case $\gamma>0$, $\gamma\ne 1$, direct calculations yield
 \begin{eqnarray*}
   III & = & (T(Q_n)-\gamma) \frac 1{\gamma^2} \int_{(t_n-u_n)/\tilde a(k_n/n)}^{(t_n+c_n-u_n)/\tilde
    a(k_n/n)}  \log(1+\gamma x) (1+\gamma x)^{-1/\gamma}\, dx
    (1+o_P(1)) \\
   & = & (T(Q_n)-\gamma) \frac 1{\gamma^2} (1+o_P(1))\times\\
   & & { }\times \frac{\big(1+\gamma
   \frac{t_n+c_n-u_n}{\tilde a(k_n/n)}\big)^{1-1/\gamma} \log\big(1+\gamma
   \frac{t_n+c_n-u_n}{\tilde a(k_n/n)}\big)-\big(1+\gamma
   \frac{t_n-u_n}{\tilde a(k_n/n)}\big)^{1-1/\gamma} \log\big(1+\gamma
   \frac{t_n-u_n}{\tilde a(k_n/n)}\big)}{\gamma-1}\\
  & = & (T(Q_n)-\gamma) \frac 1\gamma  (1+o_P(1))\times\\
  & & { } \times \frac{\big(\frac n{k_n}(1-F(t_n+c_n))\big)^{1-\gamma}
   \log\big(\frac n{k_n}(1-F(t_n+c_n))\big)-\big(\frac n{k_n}(1-F(t_n))\big)^{1-\gamma}
   \log\big(\frac n{k_n}(1-F(t_n+c_n))\big)}{1-\gamma} \\
  & = & (T(Q_n)-\gamma) \frac 1{\gamma} \Big(\frac
  n{k_n}(1-F(t_n))\Big)^{1-\gamma} \Big|\log\Big(\frac
  n{k_n}(1-F(t_n))\Big) \Big| \frac{1-\lambda^{1-\gamma}}{1-\gamma}(1+o_P(1))
 \end{eqnarray*}
 where in the last but one step again \eqref{eq:tnasymp} has been
 used. For $\gamma=1$, analogous arguments yield
 \begin{eqnarray*}
   III & = & (T(Q_n)-\gamma)  \int_{(t_n-u_n)/\tilde a(k_n/n)}^{(t_n+c_n-u_n)/\tilde
    a(k_n/n)} \frac{\log(1+x)}{1+x}-\frac{x}{(1+x)^2}\, dx
    (1+o_P(1)) \\
   & = & (T(Q_n)-\gamma) \frac 12 \bigg(\log^2\Big(\Big(\frac
   n{k_n}(1-F(t_n+c_n))\Big)^{-1}(1+o_P(1))\Big)\\
   & & { } \hspace{2.5cm} - \log^2\Big(\Big(\frac
   n{k_n}(1-F(t_n))\Big)^{-1}(1+o_P(1))\Big)+o\Big(\log\Big(\frac
   n{k_n}(1-F(t_n))\Big)\Big)\bigg)\\
   & = & (T(Q_n)-\gamma) \log\Big(\frac
   n{k_n}(1-F(t_n))\Big) (\log\lambda+o_P(1)).
 \end{eqnarray*}

 In the case $\gamma=0$ we have $\tilde\gamma_n x^2=O_P(k_n^{-1/2}
 \log^2(n(1-F(t_n))/k_n)=o_p(1)$ uniformly over the range of
 integration. A Taylor expansion of $\log(1+t)$ and of $\e^{-t}$ at
 $t=0$ shows that the integrand of III equals
 $$ \frac 1{\tilde\gamma_n^2} \log(1+\tilde\gamma_n x)\exp\Big(-\frac 1{\tilde\gamma_n}
 \log(1+\tilde\gamma_n x)\Big)-\frac
  x{\tilde\gamma_n}\exp\Big(-\Big(1+\frac 1{\tilde\gamma_n}\Big)\log(1+\tilde\gamma_n
  x)\Big)= \frac 12 x^2\e^{-x}+O_P(k_n^{-1/2} x^5\e^{-x}).
 $$
 Hence, in view of \eqref{eq:tnasymp},
 \begin{eqnarray*}
  III & = & (T(Q_n)-\gamma) \frac 12\bigg[
  \Big(\frac{t_n-u_n}{\tilde a(k_n/n)}\Big)^2 \exp\Big(-\frac{t_n-u_n}{\tilde
  a(k_n/n)}\Big)-\Big(\frac{t_n+c_n-u_n}{\tilde a(k_n/n)}\Big)^2 \exp\Big(-\frac{t_n+c_n-u_n}{\tilde
  a(k_n/n)}\Big)\\
  & & { } + O\Big(\frac{t_n-u_n}{\tilde a(k_n/n)}\exp\Big(-\frac{t_n-u_n}{\tilde
  a(k_n/n)}\Big)\Big) +  O\Big(k_n^{-1/2}\Big(\frac{t_n-u_n}{\tilde a(k_n/n)}\Big)^5\exp\Big(-\frac{t_n-u_n}{\tilde
  a(k_n/n)}\Big)\Big)\bigg]\\
  & = & (T(Q_n)-\gamma) \frac 12\bigg[
  \log ^2 \Big(\frac n{k_n}(1-F(t_n))\Big) \frac n{k_n}(1-F(t_n)) -
  \log ^2 \Big(\frac n{k_n}(1-F(t_n+c_n))\Big) \frac n{k_n}(1-F(t_n+c_n)) \\
  & & { }
  +O\Big(\log \Big(\frac n{k_n}(1-F(t_n))\Big) \frac n{k_n}(1-F(t_n))\Big)+
   O\Big(k_n^{-1/2}\log^5 \Big(\frac n{k_n}(1-F(t_n))\Big) \frac n{k_n}(1-F(t_n))\Big) \bigg]\\
   & = & (T(Q_n)-\gamma) \frac 12  \log ^2 \Big(\frac n{k_n}(1-F(t_n))\Big) \frac n{k_n}(1-F(t_n))
   (1-\lambda+o_P(1))
  \end{eqnarray*}
  by assumption \eqref{eq:kncond3}.

  To sum up, in all cases we have proved that
  $$ III = \Big(\frac n{k_n}(1-F(t_n))\Big)^{1-\gamma} \tau_n
  \frac{1-\lambda^{1-\gamma}}{1-\gamma} (1+o_P(1)).
  $$
  In view of the asymptotic normality of $T(Q_n)$, the assertion is
  obvious if we can show that the other terms in the above
  decomposition of the estimation error are of smaller order.

  To derive an upper bound for the term $I_b$, note that by \eqref{eq:Qnconv}, \eqref{eq:scaleestas} and
  \eqref{eq:tnasymp}
  \begin{eqnarray*}
    \frac{t_n-Q_n(1)}{S(Q_n)} - \frac{t_n-u_n}{\tilde a(k_n/n)} & =
    & -\Big(\frac{t_n-u_n}{\tilde a(k_n/n)}\Big(\frac{S(Q_n)}{\tilde
    a(k_n/n)}-1\Big)+ \frac{Q_n(1)-u_n}{\tilde a(k_n/n)}\Big)\frac{\tilde
    a(k_n/n)}{S(Q_n)} \\
    & = & O_P\Big(
    k_n^{-1/2}\frac{(n(1-F(t_n))/k_n)^{-\gamma}-1}\gamma\Big).
  \end{eqnarray*}
  Hence, again using \eqref{eq:tnasymp} we obtain
  $$ |I_b| = O_P\Big(    k_n^{-1/2}\frac{(n(1-F(t_n))/k_n)^{-\gamma}-1}\gamma\frac
  n{k_n}(1-F(t_n))\Big),
  $$
  which is of smaller order than the term III. Likewise, it can be
  shown that $I_a$ is asymptotically negligible.

  Next, check that by similar arguments
  $$ II = O_P(k_n^{-1/2}) \Big( \int_{(t_n-u_n)/\tilde a(k_n/n)}^{(t_n+c_n-u_n)/\tilde
    a(k_n/n)}  (1+\gamma x)^{-1/\gamma}\, dx + III\Big) = O_P\Big(
  k_n^{-1/2}\frac{(n(1-F(t_n))/k_n)^{1-\gamma}-1}{1-\gamma}\Big),
  $$
  which also is of smaller order than the term III.

  It remains to be shown that the last term IV is asymptotically
  negligible, too. For $y=u_n+\tilde a(u_n)x$ equation
  \eqref{eq:tnasymp} reads as
  $$ x= \frac{\big(\frac n{k_n}(1-F(u_n+\tilde
  a(u_n)x))\big)^{-\gamma}-1}\gamma + o\big(k_n^{-1/2} \tau_n (1+\gamma
  x)\big)
  $$
  and thus
  $$ \frac n{k_n}(1-F(u_n+\tilde  a(u_n)x)) = (1+\gamma
  x)^{-1/\gamma}\big(1+o(k_n^{-1/2}\tau_n))^{-1/\gamma} = (1+\gamma
  x)^{-1/\gamma}\big(1+o(k_n^{-1/2}\tau_n)).
  $$
  We may conclude that
  \begin{eqnarray*}
    |IV| & \le &  \int_{(t_n-u_n)/\tilde a(k_n/n)}^{(t_n+c_n-u_n)/\tilde
    a(k_n/n)} \Big| \frac n{k_n}(1-F(u_n+\tilde
    a(k_n/n)x))-(1+\gamma x)^{-1/\gamma}\Big|\, dx\\
    & = & o(k_n^{-1/2}\tau_n) \int_{(t_n-u_n)/\tilde a(k_n/n)}^{(t_n+c_n-u_n)/\tilde
    a(k_n/n)} (1+\gamma x)^{-1/\gamma}\, dx
  \end{eqnarray*}
  which is asymptotically negligible compared with III.
\end{proofof}

\begin{proofof} Corollary \ref{cor:qqapprox}. \rm
By Theorem \ref{theo:Qnlimit} (with $\tilde a(t)=\gamma \qF(1-t)$)
and Skorohod's theorem, there exist versions of $Q_n$ and $W$ such
that
$$ \sup_{0<t\le 1} t^{\gamma+1/2+\eps/2} \Big| k_n^{1/2}\Big(
\frac{Q_n(t)}{\qF(1-k_n/n)}-t^{-\gamma}\Big) -\gamma t^{-(\gamma+1)}
W(t)\Big| \;\to\; 0 \quad\text{a.s.}
$$
A Taylor expansion of the logarithm at 1 yields
\begin{eqnarray*}
  \log\Big(\frac{Q_n(T)}{Q_n(1)} t^\gamma \Big) & = & \log\big( 1+\gamma
  k_n^{-1/2} t^{-1}W(t)+o(k_n^{-1/2}t^{-(1/2+\eps)})\big) - \log\big(1+\gamma k_n^{-1/2} W(1)+o(k_n^{-1/2})\big)\\
  & = & \gamma k_n^{-1/2}(t^{-1}W(t)-W(1))+ o(k_n^{-1/2}t^{-(1/2+\eps)})
\end{eqnarray*}
uniformly for $t\in [k_n^{-1/(1+\eps)},1]$, because then $k_n^{-1/2}
t^{-1}W(t)\to 0$ uniformly by the law of the iterated logarithm for
Brownian motions. Hence
\begin{eqnarray*}
 \lefteqn{t^{1/2+\eps}\bigg[ k_n^{1/2}\Big( \log
 \frac{Q_n(t)}{Q_n(1)} + T(Q_n)\log t\Big) - \gamma
 (t^{-1}W(t)-W(1))- \int_{(0,1]} s^{-(\gamma+1)}
 W(s)\,\nu_{T,\gamma}(ds)\cdot \log t\bigg]} \\
 & = & t^{1/2+\eps}\Big(k_n^{1/2}(T(Q_n)-\gamma) -\int_{(0,1]} s^{-(\gamma+1)}
 W(s)\,\nu_{T,\gamma}(ds)\Big)\log t+o(1) \hspace*{5cm}\\
 & \stackrel{P}{\to} & 0
\end{eqnarray*}
uniformly for $t\in [k_n^{-1/(1+\eps)},1]$ by \eqref{eq:TQnconv}.

To deal with small values of $t$, recall the following well-known
facts about the order statistics of iid rv's $U_i$ that are
uniformly distributed over $(0,1)$:
\begin{equation} \label{eq:unifquantbd}
 \sup_{0<t\le 1} \frac t{nU_{[nt]+1:n}}=O_P(1), \quad
\sup_{1/(2n)\le t\le 1} \frac {nU_{[nt]+1:n}}t=O_P(1),
\end{equation}
 and $n U_{k_n+1:n}/k_n\to 1$ in probability (see, e.g., Shorack and Wellner, 1986, (10.3.7) and (10.3.8)). Because
 $(X_{n-i+1:n})_{1\le i\le k_n+1}$ has the same distribution as
 $(\qF(1-U_{i:n}))_{1\le i\le k_n+1}$, it follows by assumption
 \eqref{eq:kncond2} that for
  \begin{equation} \label{eq:Rtildedef}
 \tilde R(t,x):=\frac{\qF(1-tx)}{\qF(1-t)}-x^{-\gamma}
 \end{equation}
 one has
 \begin{eqnarray*}
  \log\Big(\frac{Q_n(t)}{Q_n(1)} t^\gamma\Big) & =^d &
  \log\Big(\frac{\qF(1-U_{[k_nt]+1:n})}{\qF(1-k_n/n)}
  t^\gamma\Big) -
  \log\frac{\qF(1-U_{k_n+1:n})}{\qF(1-k_n/n)} \\
  & = & \log\Big(\frac{nU_{[k_nt]+1:n}}{k_nt}\Big)^{-\gamma} +
  \log\Big(1+\Big(\frac{nU_{[k_nt]+1:n}}{k_n}\Big)^\gamma \tilde
  R\Big(\frac{k_n}n,\frac{nU_{[k_nt]+1:n}}{k_n}\Big)\Big) \\
  & & { } - \log\Big(\frac{nU_{k_n+1:n}}{k_n}\Big)^{-\gamma} -
  \log\Big(1+\Big(\frac{nU_{k_n+1:n}}{k_n}\Big)^\gamma \tilde
  R\Big(\frac{k_n}n,\frac{nU_{k_n+1:n}}{k_n}\Big)\Big) \\
  & = & -\gamma \log \frac{nU_{[k_nt]+1:n}}{k_nt} +
  \log\Big(1+o_P\Big(
  k_n^{-1/2}\Big(\frac{nU_{[k_nt]+1:n}}{k_n}\Big)^{-1/2}\Big)\Big)\\
  & & { }
  + \gamma \log \frac{nU_{k_n+1:n}}{k_n}-
  \log\Big(1+o_P\Big(
  k_n^{-1/2}\Big(\frac{nU_{k_n+1:n}}{k_n}\Big)^{-1/2}\Big)\Big)\\
  & = & O_P(1)
 \end{eqnarray*}
 uniformly for $t\in [(2k_n)^{-1},k_n^{-1/(1+\eps)}]$, where in
 the last step we have used \eqref{eq:unifquantbd} which implies
 $$k_n^{-1/2}\Big(\frac{nU_{[k_nt]+1:n}}{k_n}\Big)^{-1/2} \le
 (nU_{1:n})^{-1/2}=O_P(1).
 $$
Therefore,
\begin{eqnarray*}
  t^{1/2+\eps} k_n^{1/2}\Big( \log \frac{Q_n(t)}{Q_n(1)} + T(Q_n)\log t\Big)
    & = & O_P(t^{1/2+\eps}k_n^{1/2})+ k_n^{1/2}(T(Q_n)-\gamma)t^{1/2+\eps}\log
    t \\
    & = & O_P\big(k_n^{1/2-(1/2+\eps)/(1+\eps)}\big)+o_P(1) \\
    & \to & 0
\end{eqnarray*}
in probability uniformly for $t\in [(2k_n)^{-1},k_n^{-1/(1+\eps)}]$.
Now assertion \eqref{eq:qqapprox0} is obvious from the law of
iterated logarithm for Brownian motions and the fact that $Q_n$ is
constant on $(0,1/k_n)$.

If $h(t)t^{1/2+\eps}$ is bounded, the continuous mapping theorem
yields
\begin{eqnarray}  \label{eq:qqplotconv}
    \lefteqn{k_n^{1/2} \sup_{0<t\le 1} h(t) \Big( \log \frac{Q_n(t)}{Q_n(1)} + T(Q_n)\log
  t\Big)}\nonumber\\
   &\longrightarrow & \sup_{0<t\le 1} h(t)\Big(
  \gamma(t^{-1}W(t)-W(1)) + \int_{(0,1]}
  s^{-(\gamma+1)}W(s)\,\nu_{T,\gamma}(ds)\cdot\log t\Big).
\end{eqnarray}
Since $Q_n$ is constant on intervals of the form
$[(i-1)/k_n,i/k_n)$,  the continuity of $h$ implies
\begin{eqnarray*}
  \lefteqn{\max_{[k_nt_0] \le i\le k_n} h\Big(\frac{i-1/2}{k_n+1/2}\Big) k_n^{1/2}\Big|
    \log \frac{X_{n-i+1:n}}{X_{n-k_n:n}} + T(Q_n)\log
    \frac{i-1/2}{k_n+1/2}\Big|}\\
    & = & \sup_{t_0\le t\le 1}  h(t)  k_n^{1/2} \Big| \log \frac{Q_n(t)}{Q_n(1)} + T(Q_n)\log
  t\Big| +o_P(1)
\end{eqnarray*}
for all $t_0\in (0,1]$. Finally, by the law of iterated logarithm
combined with \eqref{eq:qqplotconv}
\begin{eqnarray*}
  \lefteqn{\max_{1 \le i< [k_nt_0]} h\Big(\frac{i-1/2}{k_n+1/2}\Big) k_n^{1/2} \Big|
    \log \frac{X_{n-i+1:n}}{X_{n-k_n:n}} + T(Q_n)\log
    \frac{i-1/2}{k_n+1/2}\Big|}\\
     & \le & \sup_{0<t\le t_0}  h(t)  k_n^{1/2} \Big| \log \frac{Q_n(t)}{Q_n(1)} + T(Q_n)\log
  t\Big| +o_P(1) \\
  & \to & 0
\end{eqnarray*}
in probability as $t_0\to 0$, so that assertion \eqref{eq:qqapprox1}
follows. The last assertion is immediate from
$\nu_{T_H,\gamma}(ds)=\gamma(s^\gamma ds-\eps_1(ds))$ with $\eps_1$
denoting the Dirac measure in 1 (cf.\ Drees (1998b), Example 3.1).
\end{proofof}

\begin{proofof} Corollary \ref{cor:qqapprox1}. \rm
  Note that
  \begin{equation}  \label{eq:Rtilderep}
    \frac{\qF(1-tx)}{\qF(1-t)}x^\gamma = \exp\Big(\int_{tx}^t
    \frac{\eta(s)}s\, ds\Big),
  \end{equation}
   so that condition
  \eqref{eq:kncond2} reads as
  \begin{eqnarray*}
   \lefteqn{k_n^{1/2} \sup_{0<x\le 1+\tilde\eps}x^{1/2}
  \Big(\exp\Big(\int_x^1\frac{\eta(sk_n/n)}s\, ds\Big)-1\Big|}\\
  & \le & k_n^{1/2} \sup_{0<x\le 1+\tilde\eps}x^{1/2}
  \Big|\exp\Big(\sup_{0<s\le (1+\tilde\eps)k_n/n}|\eta(s)|\log
  x\Big)-1\Big|\\
  & \to & 0.
  \end{eqnarray*}
  In view of \eqref{eq:kncond6} this condition is fulfilled, and so convergence
  \eqref{eq:qqapprox0} holds.

  By the law of the iterated logarithm
  \begin{eqnarray*}
   h(t)\Big|\frac{W(t)}t-W(1)\Big| & \le & h(t)\frac{|W(t)-W(1)|}t +
   h(t) \frac{1-t}t |W(1)|\\
   & = & O\big( h(t)(1-t)^{1/2}\log^{1/2}|\log(1-t)|\big) +
   O(h(t)(1-t))\\
   & \to & 0 \quad a.s.
  \end{eqnarray*}
  as $t\uparrow 1$. Hence, in view of \eqref{eq:qqapprox0}, it
  suffices to prove that
  \begin{equation}  \label{eq:convpr1}
    k_n^{1/2} \sup_{t_n\le t\le 1-(2k_n)^{-1}} h(t)\Big|\log
    \frac{Q_n(t)}{Q_n(1)}+T(Q_n)\log t\Big| \;\to\; 0
  \end{equation}
  in probability for all sequences $t_n\uparrow 1$. Because
  $h(t)\log t\to 1$ as $t\uparrow 1$ and
  $k_n^{1/2}(T(Q_n)-\gamma)=O_P(1)$, \eqref{eq:convpr1} would follow
  from
  \begin{equation}  \label{eq:convpr2}
    k_n^{1/2} \sup_{t_n\le t\le 1-(2k_n)^{-1}} h(t)\Big|\log\Big(
    \frac{Q_n(t)}{Q_n(1)}t^\gamma\Big)\Big| \;\stackrel{P}{\to}\; 0.
  \end{equation}

   To establish \eqref{eq:convpr2}, one may argue similarly as in the second part of the proof of
   Corollary \ref{cor:qqapprox} using the uniform tail empirical
   quantile function, but it is easier to work with a Hungarian
   construction for partial sums $S_i:=\sum_{j=1}^i \xi_j$ with
   $\xi_j$, $j\in\N$, denoting iid standard exponential rv's. More concretely, for
   suitable versions of $\xi_j$ there exist a Brownian motion $W$
   such that $\max_{1\le i\le k_n+1}|S_i-i-W(i)|=O(\log k_n)$ a.s.
   Moreover, the variational distance between the distribution of
   $(X_{n-i+1:n})_{1\le i\le k_n}$ and the distribution of
   $(\qF(1-S_i/n))_{1\le i\le k_n}$ tends to 0 (Reiss, 1989,
   Theorem 5.4.3). Hence, to verify \eqref{eq:convpr2}, it suffices
   to prove that
   \begin{eqnarray} \label{eq:convpr3}
     \lefteqn{  k_n^{1/2} \sup_{t_n\le t\le 1-(2k_n)^{-1}} h(t)\Big|\log\Big(
    \frac{\qF(1-S_{[k_n t]+1}/n)}{\qF(1-S_{k_n+1}/n)}t^\gamma\Big)\Big| }\nonumber\\
    & \le &  k_n^{1/2} \sup_{t_n\le t\le 1-(2k_n)^{-1}}
    h(t)\gamma\Big|\log \frac{S_{[k_n t]+1}}{tS_{k_n+1}}\Big| + k_n^{1/2} \sup_{t_n\le t\le 1-(2k_n)^{-1}}
    h(t) \int_{S_{[k_n t]+1}/n}^{S_{k_n+1}/n} \frac{|\eta(s)|}s\,
    ds\\
    & \to& 0 \nonumber
   \end{eqnarray}
   in probability.

   The Hungarian construction and the law of iterated logarithm
   imply
   \begin{eqnarray*}
    \log \frac{S_{[k_n t]+1}}{S_{k_n+1}}
     & = & -\log   \frac{[k_nt]+W([k_nt]+1)+O(\log k_n)}{k_n+W(k_n+1)+O(\log
     k_n)}\\
     & = & -\log t-\log\Big(1+\frac{W([k_nt]+1)}{k_nt}+O\Big(\frac{\log
     k_n}{k_n}\Big)\Big) + \log\Big(1+\frac{W(k_n+1)}{k_n}+O\Big(\frac{\log
     k_n}{k_n}\Big)\Big) \\
     & = & 1-t +O((1-t)^2) + O\Big(\frac{\log\log
     k_n}{k_n}\Big)^{1/2}\Big).
    \end{eqnarray*}
    Thus the second term in \eqref{eq:convpr3} can be bounded by
    $$  k_n^{1/2} \sup_{0<s\le (1+\tilde\eps)k_n/n} |\eta(s)|\cdot
    \sup_{t_n\le t\le 1-(2k_n)^{-1}} h(t)\Big(1-t +O((1-t)^2) + O\Big(\Big(\frac{\log\log
     k_n}{k_n}\Big)^{1/2}\Big)\Big)
    $$
    which tends to 0 because of \eqref{eq:kncond6} and
    $h(t)(1-t)^{1/2-\eps}\to 0$ as $t\uparrow 1$.

    Moreover,
    $$ \log \frac{S_{[k_n t]+1}}{tS_{k_n+1}} =
    \log\Big(1-\frac{W([k_nt]+1)-W(k_n+1)t+O(\log
    k_n)}{k_n+O(k_n^{-1/2}\log^{1/2}\log k_n)}\Big),
    $$
    where
    \begin{eqnarray*}
      W([k_nt]+1)-W(k_n+1) & =^d & k_n^{1/2}
      W\Big(1-\frac{[k_nt]}{k_n}\Big) \\
      & = & O\Big(
      k_n^{1/2}\Big(1-\frac{[k_nt]}{k_n}\Big)^{1/2}\log^{1/2}\Big|\log\Big(1-\frac{[k_nt]}{k_n}\Big)\Big|\Big)\\
      & = & O\big(k_n^{1/2}(1-t)^{1/2} \log^{1/2}|\log(1-t)|\big)
    \end{eqnarray*}
    uniformly for $t_n\le t\le 1-(2k_n)^{-1}$. Hence, the first term
    in  \eqref{eq:convpr3} is of the stochastic order
    \begin{eqnarray*}
      \lefteqn{k_n^{1/2} \sup_{t_n\le t\le 1-(2k_n)^{-1}} h(t)\Big|
      \frac{
      W([k_nt]+1)-W(k_n+1)+W(k_n+1)(1-t)}{k_n}}\\
      \lefteqn{\hspace*{4cm}\quad +O\big(k_n^{-1}(1-t)\log|\log(1-t)|+k_n^{-1}\log^2k_n\big)\Big|} \\
      & = & O_P\Big(\sup_{t_n\le t\le 1-(2k_n)^{-1}}
      h(t)(1-t)^{1/2}\log^{1/2}|\log(1-t)|\Big)\hspace*{4cm}\\
      & \to & 0
    \end{eqnarray*}
    in probability, which completes the proof of
    \eqref{eq:qqapprox4}. The second assertion follows exactly as in
    the proof of Corollary \ref{cor:qqapprox}.
\end{proofof}

\bigskip

{\bf Acknowledgement:} I thank the anonymous referees for their
constructive criticism and remarks that led to a substantial
improvement of the presentation.
\bigskip

\noindent {\large\bf References}
   \smallskip

\parskip1.2ex plus0.2ex minus0.2ex
\def\rueck{\noindent\hangafter=1 \hangindent=1.3em}

\rueck
  Abdous, B., Foug\`{e}res, A.-L., and Ghoudi, K. (2005). Extreme
  behavior for bivariate elliptical distributions. {\em rev.\
  Canad.\ Statist.} {\bf 33}, 1095--1107.

\rueck
  Abdous, B., Foug\`{e}res, A.-L., Ghoudi, K., and Soulier, P.\ (2008). Estimation of bivariate excess probabilities
for elliptical models. {\em Bernoulli} {\bf 14}, 1065–-1088.

\rueck
  Beirlant, J., Goegebeur, Y., Segers, J., Teugels, J., de Waal, D.,
  and Ferro, C. (2004). {\em Statistics of Extremes.} Wiley.

\rueck
 Cebri\'{a}n, A.C., Denuit, M., and Lambert, P. (2003).
  Generalized Pareto fit to the society of
actuaries' large claims database. {\em N.\ Am.\ Actuar.\ J.} {\bf
7}, 18--36.

\rueck
  Charpentier, A., and Juri, A. (2006).  Limiting dependence structures for tail events, with applications to
credit derivatives. {\em J.\ Appl.\ Probab.} {\bf 43}, 563--586.

\rueck
 Charpentier, A., and Segers, J. (2009). Tails
of multivariate Archimedean copulas. {\em J.\ Multiv.\ Anal.} {\bf
100}, 1521--1537.

\rueck
  Danielsson , J., de Haan, L., Peng, L., and de Vries, C.G. (2001).
  Using a bootstrap method to choose the sample fraction in tail
  index estimation. {\em J.\ Multiv.\ Anal.} {\bf 76}, 226--248.

\rueck
  Das, B., and Resnick, S.I. (2011). Conditioning on an extreme component:
Model consistency and regular variation on cones.  {\em Bernoulli}
{\bf 17}, 226–-252.

\rueck
  Dietrich, D., de Haan, L., and H\"{u}sler, J. (2002) Testing extreme
  value conditions. {\em Extremes} {\bf 5}, 71--85.

\rueck
  Draisma, G., Drees, H., Ferreira, A., and de Haan, L. (2004).
  Bivariate tail estimation: dependence in asymptotic independence.  {\em Bernoulli} {\bf 10}, 251--280.

\rueck
  Drees, H. (1998a). On smooth statistical tail functionals.
  {\em Scand.\ J.\ Statist.} {\bf 25}, 187--210.

\rueck
  Drees, H. (1998b). A general class of estimators of
  the extreme value index. {\em J.\ Statist.\ Plann.\
  Inference} {\bf 66}, 95--112.

\rueck
  Drees, H. (2003). Extreme Quantile Estimation for Dependent
   Data with Applications to Finance.  {\em Bernoulli} {\bf 9}, 617--657.

\rueck
  Drees, H., de Haan, L., and Resnick, S. (2000).
  How to make a Hill plot. {\em Ann.\ Statist.} {\bf 28}, 254--274.

\rueck
  Drees, H., and Kaufmann, E. (1998). Selecting the optimal
   sample fraction in univariate extreme value statistics. {\em Stoch.\ Processes Appl.} {\bf 75}, 149--172.

\rueck
  Drees, H., Ferreira, A., and de Haan, L. (2004). On the maximum likelihood estimation
  of the extreme value index. {\em Ann.\ Appl.\ Probab.} {\bf 14}, 1179--1201.

\rueck
  Drees, H., and M\"{u}ller, P. (2008). Fitting and validation of a bivariate model for large claims.
  {\em Insurance Math.\ Econom.} {\bf  42}, 638--650.

\rueck
  Einmahl, J.H.J., and Segers, J. (2009). Maximum empirical
likelihood estimation of the spectral measure of an extreme value
distribution. {\em Ann.\ Statist.} {\bf 37}, 2953--2989.

\rueck
   Embrechts, P., Kl\"{u}ppelberg, C.\ and Mikosch, T. (1997).
  {\em Modelling Extremal Events}, Springer.

\rueck
  Foug\`{e}res, A.-L., and Soulier, P. (2010). Limit conditional
  distributions for bivariate vectors with polar representation.
  {\em Stoch.\ Models} {\bf 26}, 54--77.

\rueck
  Grazier, K.L., and G'Sell Associates (1997):
  Group Medical Insurance Large Claims Data Base Collection and
  Analysis.
  {\em SOA Monograph M-HB97-1}, Society of Actuaries, Schaumburg,
  Illinois.

\rueck
  de Haan, L., and Ferreira, A. (2006). {\em Extreme Value Theory.}
  Springer.

\rueck
  Ledford, A.W., und Tawn, J.A. (1996). Statistics for near
independence in multivariate extreme values. {\it Biometrika} {\bf
83}, 169--187.

\rueck
  Ledford, A.W., and Tawn, J.A. (1997).
  Modelling Dependence within Joint Tail Regions,
  {\em J. Royal Statist.\ Soci.} {\bf 59}, 475--499.

\rueck
 McNeil, A. (1997). Estimating the tails of loss severity
distributions using extreme value theory. {\em ASTIN Bull.} {\bf
27}, 117--137.

\rueck
  Mikosch, T. (2006). Copulas: Tales and facts (with discussion). {\em Extremes}. {\bf
  9}, 3--62.

\rueck
  Reiss, R.-D. (1989). {\em Approximate Distributions of Order
  Statistics.} Springer.

\rueck
  Resnick, S.I. (2007). {\em Heavy-Tail Phenomena}. Springer.

\rueck
  Schmutz, M., and Doerr, R.R. (1998).
  The Pareto model in property reinsurance. Swiss Re publication.

\rueck
  Shorack, G.R., and Wellner, J.A. (1986). {\em Empirical Processes
  with Applications to Statistics.} Wiley.

\rueck
 Vandewalle, B., and Beirlant, J. (2006). On univariate extreme value statistics and the estimation
of reinsurance premiums. {\em Insurance Math.\ Econom.} {\bf 38},
441--459.

\end{document}